\documentclass[a4paper,12pt]{article}
\usepackage{epsfig,float}
\usepackage{cite}
\usepackage[left=2cm,right=2cm]{geometry}
\usepackage{color}
\usepackage{a4wide,graphicx}
\usepackage{morefloats}
\newcommand\T{\rule{0pt}{4ex}}
\newcommand\B{\rule[-1.5ex]{0pt}{0pt}}
\newcommand{\eps}{\epsilon}

\begin{document}

\title{A new interpretation for the $D^*_{s2}(2573)$ and the prediction of novel exotic charmed mesons}

\author{R. Molina $^1$, T. Branz$^2$ and E. Oset$^1$}
\maketitle

\date{}
\begin{center}
$^1$ Departamento de F\'{\i}sica Te\'orica and IFIC,
Centro Mixto Universidad de Valencia-CSIC,
Institutos de Investigaci\'on de Paterna, Aptdo. 22085, 46071 Valencia, Spain\\
$^2$ Institut f\"ur Theoretische Physik, Universit\"at T\"ubingen, Kepler Center for Astro and Particle Physics, Auf der Morgenstelle 14, D-72076 T\"ubingen, Germany\\
\end{center}

\begin{abstract}  
In this manuscript we study the vector - vector interaction within the hidden gauge formalism in a coupled channel unitary approach. In the sector $C=1,S=1,J=2$ we get a pole in the T-matrix around $2572$ MeV that we identify with the $D^*_{s2}(2573)$, coupling strongly to the $D^*K^*$($D^*_s\phi$($\omega$)) channels. In addition we obtain  resonances in other exotic sectors which have not been studied before such as $C=1,S=-1$, $C=2,S=0$ and $C=2,S=1$. This 'flavor-exotic' states are interpreted as $D^*\bar{K^*}$, $D^*D^*$ and $D^*_sD^*$ molecular states but have not been observed yet. In total we obtain nine states with different spin, isospin, charm and strangeness of non $C=0,S=0$ and $C=1,S=0$ character, which have been reported before.

%We obtain from the model new interesting predicted states not observed yet that can be thought of as $D^*\bar{K^*}$, $D^*D^*$ and $D^*_sD^*$ molecular states correspondingly.
\end{abstract}

\section{Introduction}

The $D^*_{s2}(2573)$ was first observed by the CLEO Collaboration in 1994 \cite{Kubota}. Within the heavy quark symmetry framework (HQS) the spin of the heavy quark and the total angular momentum of the light quark are separately conserved. As a consequence, the heavy-light systems can be grouped in one doublet with $j_l=3/2$ and $J^P=1^+,2^+$ and a second doublet with $j_l=1/2$ and $J^P=0^+,1^+$. Here $j_l$ denotes the total spin of the light quark.  While the $j_l=3/2$ states are relatively narrow, the states of the $j_l=1/2$ doublet are very broad \cite{Godfrey2}. When the $D^*_{s2}(2573)$ was observed for the first time it was regarded as the possible $j_l=3/2$-doublet partner of the $D^*_{s1}(2536)$ in this picture. However, the quark model reveals some problems in the $c\bar{s}$ spectrum. First of all the doublet with $J^P=0^+,1^+$ has not been observed for a long time although it is predicted to be very broad \cite{Godfrey2}.  Further on the later discovery of the $D^*
 _{s0}(2317)$ and the $D_{s1}(2460)$ by the CLEO \cite{Besson} and BABAR \cite{Aubert} collaborations are difficult to explain in terms of quark models.

 Even if the $J^P=0^+$ assignment for the $D^*_{s0}(2317)$ meson gets confirmed, the  $D^*_{s0}(2317)$ and $D_{s1}(2460)$ masses and widths are in contradiction to typical quark model predictions. The physical masses lie around 100 MeV below the quark potential model which estimates a mass of the $D^*_{s0}(2317)$ of $2.48$ \cite{Godfrey2,Godfrey1} or $2.49$ GeV \cite{Pierro} and $2.53-2.57$ GeV for the $D_{s1}(2460)$.

 %   , there appears to be a discrepancy with typical quark potential models which predict a mass of $2.48$ \cite{Godfrey2,Godfrey1}, or $2.49$ GeV \cite{Pierro}. Whereas in the case of the $D_{s1}(2460)$, with $J^P=1^+$, the predicted masses are $2.53$ and $2.57$ GeV for the $1^1 P_1$ and $1^3 P_1$ states respectively in the $c\bar{s}$ spectrum \cite{Godfrey1}. 
%These masses are about $100$ MeV over the real mass. 
In addition, the widths for these two states are very small, $<3.5$ and $<3.8$ MeV for the $D^*_{s0}(2317)$ and $D_{s1}(2460)$ respectively.  This is in disagreement with the HQS prediction  expecting a broad $j_l=1/2$ doublet with $J^P=0^+,1^+$. A possible solution suggested by many authors is that the strong S-wave coupling of the $D^*_{sJ}$ states to the $DK$($D^*K$) decay channel and the proximity to the thresholds  could shift the respective masses \cite{vanBeveren,Hwang,Simonov,Becirevic,Eichten,Rupp}. %Therefore, the "quark-antiquark" picture is not adequate particularly in the case of the $c\bar{s}$ spectrum and other pictures of 4-quarks could be more likely \cite{vanBeveren} being the strong S-wave coupling to PP and PV the key to the unusual properties of the new light $D_{sJ}$ mesons. 
 Since the standard $c\bar{s}$ scenario is in disagreement with experimental observations, alternative structure interpretations have been made.  For instance a 4-quark picture could be more likely \cite{vanBeveren},  where the strong S-wave coupling to PP and PV  might be the key to the unusual properties of the new light $D_{sJ}$ mesons.

In \cite{Gamermann1} two different models are used  to study coupled channels of pseudoscalar mesons. In the first approach  the PP interaction is set up by a chiral Lagrangian  while the second method is provided by a phenomenological model based on a $SU(4)$ symmetric Lagrangian. Subsequently, the symmetry is broken down to SU(3) by identifying the suppressed currents where heavy vector-mesons are exchanged. Both models, the chiral Lagrangian and the phenomenological model, lead to very similar results. The unitarization in the coupled channel formalism generates dynamically the $D_{s0}(2317)$ as a bound state from the $DK$ and $D_s\eta$ channels essentially. Here, the chiral symmetry can be restored by setting this new SU(4) symmetry breaking parameters to zero and using a unique $f_\pi$ parameter \cite{medio}. The results of \cite{Gamermann1} are comparable to those obtained in \cite{Faessler:2007gv} where a effective Lagrangian approach is used assuming a pure $DK$ molec
 ular structure for the $D_{s0}(2317)$. In addition, similar results are obtained in \cite{lutz,guo1} omitting the exchange of heavy vector mesons. In a later work the coupled channel analysis of \cite{Gamermann1} was extended by a phenomenological model for the PV interaction \cite{Gamermann2}.  As a conclusion, the $D_{s1}(2460)$ and the $D_{s1}(2536)$ are obtained in this work as very narrow peaks from the $KD^*$($\eta D^*_s$) and $D K^*$($D_s\omega$($\phi$)) channels, respectively. We emphasize that in this work very few parameters are used in comparison with the large amount of information obtained. Similarly, in \cite{Faessler:2007us,guo2} the $D_{s1}(2460)$ is also considered as a hadronic bound state of a $K$ and a $D^\ast$ meson. The
work of \cite{wise} used a chiral Lagrangian based on heavy quark
symmetry for the open charm sector which neglects
exchanges of heavy vector mesons in the implicit
Weinberg-Tomozawa term.

 The success of the $PP$ and $PV$ coupled channels in the charm-strange sector motivates the extension to $VV$ interaction. The vector meson interaction can be included in the chiral Lagrangian by means of the hidden gauge formalism. In the present paper we extend the two meson molecular idea to two vector mesons. We concentrate on dynamically generated resonances with charm-strange ($C=1,S=1$) quantum numbers and exotic sectors which have not been addressed before from this point of view.
 
% we elaborate further on the two meson molecular idea extending it to two vector mesons. We concentrate, however, in the $C=1,S=1$, and some other sectors, which have not been addressed before from this point of view.
%How the VV interaction is implied in the nature of some of these resonances with $C=1;S=1$, in particular (and also in other sectors), this is the first time this point of view is taken and what we try to understand in this work. 
%Previous works about the possible VV nature of some resonances has been done. 
In \cite{raquel1} the authors used the hidden gauge lagrangian, together with a unitary approach, to study the $\rho\rho$ interaction. 
The potential was strong enough to bind the $\rho\rho$ system and two states around $1270$ and $1500$ MeV were obtained as poles in the $\rho\rho$ scattering amplitude. They were identified with the $f_2(1270)$ and $f_0(1375)$ respectively. The decay of these resonances was provided by means of box diagrams with two or four pions in the intermediate state. This mechanism provided a width of the order of $110$ and $200$ MeV respectively for these states, which is comparable to the data in the PDG \cite{pdg}. Actually, there are strong experimental arguments which support the $\rho \rho$ molecule interpretation of the  $f_0(1370)$ \cite{klempt,crede}. In \cite{geng} the authors extended the model to SU(3) resulting in eleven poles in the scattering matrix, bound states or resonances. Five of them can be identified with states quoted in the PDG: $f_0(1370)$, $f_0(1710)$, $f_2(1270)$, $f_2'(1535)$ and $K^*_2(1430)$ (see Table IV of \cite{geng}). 
 The analysis of processes involving these states further support their interpretation as dynamically generated states. In this direction, the radiative decay of the $f_0(1370)$ and $f_2(1270)$ mesons into $\gamma \gamma$ was calculated in \cite{yamagata}, where the authors found a good agreement with the experimental data. Similarly the $J/\psi$ decay into $\phi (\omega)$ and one of the $f_2(1270)$, $f'_2(1525)$, $f_0(1710)$ resonances as well as the process $J/\psi\to K^*K^*_2(1430)$ was also found to be consistent with experiment \cite{chinacola}. In the same line, the  $J/\psi$ radiative decay into
$\gamma$ and one of these non-strange resonances was also able to reproduce experimental data \cite{chinavalgerman}. Recently, the $\gamma\gamma$ and $\gamma$-vector meson decays of the eleven dynamically generated resonances  of \cite{geng} have been studied in  \cite{BranzGeng} and the decay widths are in good agreement with data where these are available.

%Later works \cite{raquel2,xyz} have extended the model followed in \cite{raquel1,geng} to channels like $\rho D^*$ or $D^*_{(s)}\bar{D}^*_{(s)}$ involving mesons with a $c$-quark in a similar way as it was done with the PP or as PV channels involving the $D$ or $D^*$ meson in \cite{Gamermann1,Gamermann2}. 
The model applied in \cite{raquel1,geng} was in later works extended to channels with a charmed meson involved \cite{raquel2,xyz}. The authors proceeded in a similar way as in case of the inclusion of $D$ and $D^\ast$ mesons in the  PP or as PV channels \cite{Gamermann1,Gamermann2}.
A SU(4)-symmetric lagrangian for the three and four-vector interaction is constructed and once one builds the vector-exchange diagrams the symmetry is broken by suppressing those terms where a heavy vector meson is exchanged. In \cite{raquel2} the attraction between the $\rho$($\omega$) meson and the $D^*$ is strong enough to bind the $\rho$($\omega$)$D^*$ system and three states are obtained for $I=1/2$ and $J^P=0^+,1^+,2^+$ respectively: the $D_0(2600)$, $D^*(2640)$ and $D^*_2(2460)$. The first one, with a width around $61$ MeV is a prediction of the model and the third state appears naturally in the scheme. The $D^*(2640)$ is obtained with a small width of $3-4$ MeV since the decay to two pseudoscalar mesons ($\pi D$) by means of a box diagram is forbidden for the quantum numbers $J^P=1^+$. In particular this small width is the main reason to associate the $D^*(2640)$ to the $J^P=1^+$ quantum numbers. Therefore, one finds a reasonable explanation on why the $D^*(2640)$ is 
 a very narrow state in comparison with the $D^*_2(2460)$, even though the first one has a larger mass. In \cite{xyz} the authors study the region of $4000$ MeV with a set of $16$ channels for $C=0$, $S=0$ and $I=0$ or $1$. They obtained five poles in the scattering matrix, three of which could be identified by the proximity of the mass, width and quantum numbers with the Y(3940), Z(3930) and X(4160), corresponding to hadronic molecules made of $D^*\bar{D}^*$, $D_s^*\bar{D}_s^*$. The radiative decay of these resonances in $PV\gamma$ was studied in \cite{weihong}. The strong hidden charm decay mode $J/\psi \omega$ and the two-photon decay of the $Y(3940)$ within a $D^*\bar{D}^*$ bound state interpretation is also discussed in \cite{Branz:2010qw}.

In the present work we follow the same approach as in \cite{raquel2,xyz} in order to study $VV$ coupled channels in the hidden-charm ($C=0;S=1$) and charm-strange sector ($C=1;S=-1$). Further on we also extend our formalism to 'flavor-exotic' channels as for instance $C=1;S=1$, $C=1;S=2$, $C=2;S=0$, $C=2;S=1$ and $C=2;S=2$.

%we are going to follow the same approach than in \cite{raquel2,xyz} to study the VV interaction in the sectors that were not been studied there: $C=0;S=1$ (hidden charm), $C=1;S=-1$, $C=1;S=1$, $C=1;S=2$, $C=2;S=0$, $C=2;S=1$ and $C=2;S=2$.

\section{Formalism}

The hidden-gauge formalism is applied in order to describe the interaction between vector mesons and vector mesons with pseudoscalars and photons \cite{hidden1,hidden2,hidden3,hidden4}. The hidden-gauge Lagrangian, which is consistent with chiral symmetry, provides this former interaction from the following terms
\begin{equation}\label{eq:laghg}
\mathcal{L}=-\frac{1}{4}\langle \bar{V}_{\mu\nu}\bar{V}^{\mu\nu}\rangle
+\frac{1}{2}M_v^2 \langle [V_\mu-(i/g)\Gamma_\mu]^2\rangle,
\end{equation}
where
\begin{equation}
\bar{V}_{\mu\nu}=\partial_\mu V_\nu-\partial_\nu V_\mu-ig[V_\mu,V_\nu],
\end{equation}
\begin{equation}
\Gamma_\mu=\frac{1}{2}\left\{
u^\dagger [\partial_\mu -i(v_\mu+a_\mu)]u +
u[\partial_\mu-i(v_\mu-a_\mu)]u^\dagger\right\},
\end{equation}
and $\langle\rangle$ stands for the trace in the SU(3) flavor space.
$V_\mu$ represents the vector nonet:
 \begin{equation}
\renewcommand{\tabcolsep}{1cm}
\renewcommand{\arraystretch}{2}
V_\mu=\left(
\begin{array}{ccc}
\frac{\omega+\rho^0}{\sqrt{2}} & \rho^+ & K^{*+}\\
\rho^- &\frac{\omega-\rho^0}{\sqrt{2}} & K^{*0}\\
K^{*-} & \bar{K}^{*0} &\phi
\end{array}
\right)_\mu,
\end{equation}
where $u^2=U=\exp\left(\frac{i\sqrt{2}\Phi}{f}\right)$ and
$\Phi$ is the octet of the pseudoscalars
\begin{equation}
\renewcommand{\tabcolsep}{1cm}
\renewcommand{\arraystretch}{2}
\Phi=\left(
\begin{array}{ccc}
\frac{\eta}{\sqrt{6}}+\frac{\pi^0}{\sqrt{2}} & \pi^+ & K^+\\
\pi^- &\frac{\eta}{\sqrt{6}}-\frac{\pi^0}{\sqrt{2}} & K^{0}\\
K^{-} & \bar{K}^{0} &-\sqrt{\frac{2}{3}}\eta
\end{array}
\right).
\end{equation}
The use of the value of the coupling constant $g$ of the Lagrangian (Eq.~(\ref{eq:laghg}))
is given by
\begin{equation}
g=\frac{M_V}{2f},
\label{eq:g}
\end{equation}
with the pion decay constant $f=93$ MeV. The use of the value of $g$ of Eq. (\ref{eq:g}) provides one way to account for the Kawarabayashi-Suzuki-Fayyazuddin-Riazuddin (KSFR) relation \cite{KSFR}, which is tied to the vector meson dominance  formalism \cite{sakurai}. In order to incorporate the charmed mesons we do a straightforward extension of the $V_\mu$ matrix to SU(4), as it was done in \cite{Gamermann1,Gamermann2,raquel2,xyz}:
\begin{equation}
\renewcommand{\tabcolsep}{1cm}
\renewcommand{\arraystretch}{2}
V_\mu=\left(
\begin{array}{cccc}
\frac{\omega+\rho^0}{\sqrt{2}} & \rho^+ & K^{*+}&\bar{D}^{*0}\\
\rho^- &\frac{\omega-\rho^0}{\sqrt{2}} & K^{*0}&D^{*-}\\
K^{*-} & \bar{K}^{*0} &\phi&D^{*-}_s\\
D^{*0}&D^{*+}&D^{*+}_s&J/\psi\\
\end{array}
\right)_\mu.
\end{equation}
\begin{figure}
\begin{center}
\includegraphics[width=16cm]{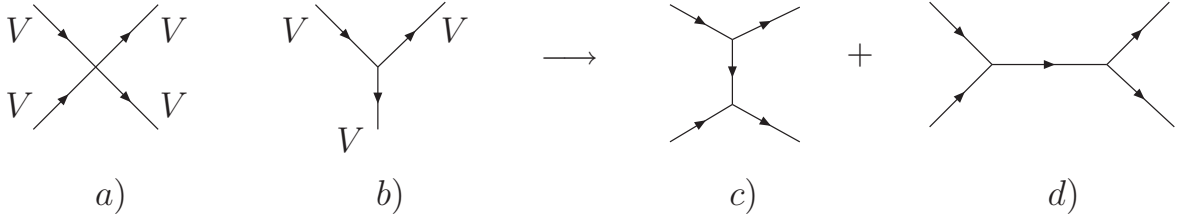}
\end{center}
\caption{Terms of the ${\cal L}_{III}$ Lagrangian: a) four-vector contact term,
 Eq.~(\ref{eq:4v}); b) three-vector interaction, Eq.~(\ref{eq:3v}); c) $t$ and 
 $u$ channels from vector exchange; d) $s$ channel for vector exchange.}
\label{fig:fig1} 
\end{figure}
 Subsequently, the symmetry is broken down by taking the heavy masses of the charmed mesons into account, in particular suppressing the heavy-meson exchange terms. The Lagrangian of Eq. (\ref{eq:laghg}) provides the four-vector and three-vector contact terms
\begin{equation}\label{eq:4v}
\mathcal{L}_\mathrm{VVVV}=\frac{1}{2}g^2\langle [V_\mu,V_\nu]V^\mu V^\nu\rangle,
\end{equation}
\begin{eqnarray}\label{eq:3v}
\mathcal{L}_{VVV}&=&ig\langle (\partial_\mu V_\nu-\partial_\nu
V_\mu)V^\mu V^\nu\rangle\nonumber\\
&=&ig\langle V^\mu \partial_\nu V_\mu V^\nu-\partial_\nu V_\mu V^\mu
V^\nu \rangle\nonumber\\
&=&ig\langle (V^\mu \partial_\nu V_\mu -\partial_\nu V_\mu V^\mu)
V^\nu)\rangle.
\end{eqnarray}
The three-vector contact terms lead to the vector-exchange diagrams of Figs. \ref{fig:fig1} c) and d). In the approximation of low momenta of the external vectors compared to the mass of the vector mesons, $\vec{k}/M_V\sim 0$, the polarization vectors of the external vector mesons reduce to the spatial components. This implies that the vector field $V^\nu$ in Eq. (\ref{eq:3v}) corresponds necessarily to the exchanged vector meson. Indeed, if it were an
 external vector meson, the $\nu$ index should be spatial as already mentioned. Then, the derivative $\partial_\nu$ would lead to a three-momentum of an external
 vector or a difference of two of them, which are neglected in the present approach. Eq. (\ref{eq:3v}) leads to the amplitudes of the diagram of Fig. \ref{fig:fig1} c) ($V_1(k_1)V_2(k_2)\to V_3(k_3)V_4(k_4)$) which in the $t$-channel reads as
 \begin{equation}\label{eq:tch}
 (k_1+k_3)\cdot (k_2+k_4)\;\epsilon_1\cdot\epsilon_3
 \,\epsilon_2\cdot\epsilon_4,
 \end{equation}
 whereas the amplitudes corresponding to $u$-channel diagrams are of the type
 \begin{equation}\label{eq:uch}
 (k_1+k_4)\cdot(k_2+k_3)\;\epsilon_1\cdot\epsilon_4 \,
 \epsilon_2\cdot\epsilon_3.
 \end{equation}
  In general, the diagrams in the s-channel (see Fig. \ref{fig:fig1} d)) are also possible. However,  according to \cite{raquel1} these amplitudes lead to a repulsive p-wave interaction for equal masses of the vectors and only to a minor s-wave component in the case of different masses \cite{geng}. Therefore, we can neglect the diagrams of Fig. \ref{fig:fig1} d) completely.
 
 By neglecting the three-momenta of the external vector mesons with respect to the mass, only the spatial components of the polarization vectors remain, and one can easily build the spin-projection operators \cite{raquel1}, which are 
  \begin{eqnarray}
{\cal P}^{(0)}&=& \frac{1}{3}\eps_\mu \eps^\mu \eps_\nu \eps^\nu\nonumber\\
{\cal P}^{(1)}&=&\frac{1}{2}(\eps_\mu\eps_\nu\eps^\mu\eps^\nu-\eps_\mu\eps_\nu\eps^\nu\eps^\mu)\nonumber\\
{\cal P}^{(2)}&=&\lbrace\frac{1}{2}(\eps_\mu\eps_\nu\eps^\mu\eps^\nu+\eps_\mu\eps_\nu\eps^\nu\eps^\mu)-\frac{1}{3}\eps_\mu\eps^\mu\epsilon_\nu\epsilon^\nu\rbrace\ .
\label{eq:projmu}
\end{eqnarray}

Thus, the spin projections of the structures of Eq. (\ref{eq:tch}) and Eq. (\ref{eq:uch}) can be written as
\begin{eqnarray}
(k_1+k_3)\cdot (k_2+k_4)\hspace{0.3cm} \mathrm{for}\hspace{0.3cm} J=0,1,2\ ,
\end{eqnarray}
and 
\begin{eqnarray}
& &(k_1+k_4)\cdot(k_2+k_3)\hspace{0.3cm} \mathrm{for}\hspace{0.3cm} J=0,2\ ,\nonumber\\
&-&(k_1+k_4)\cdot(k_2+k_3)\hspace{0.3cm} \mathrm{for}\hspace{0.3cm} J=1\ ,
\end{eqnarray}
respectively. The tree-level transition amplitudes from the four-vector contact terms and vector-exchange terms are listed in the Appendix. The value of $g$ in these tables is set to $g=M_\rho/2\,f_\pi$. As one can observe from these tables, the potential from the four-vector contact terms plus vector-exchange diagrams lead to a strong attractive interaction for the quantum numbers: $C=1,S=-1,I=0,J=0,1,2$; $C=1,S=1,I=0,1,J=0,1,2$; $C=2, S=0,I=0,J=1$ and $C=2,S=1,I=1/2,J=1$, whereas we obtain repulsion or a very small contribution ($\sim g^2$) in the other sectors. This is in addition to the $C=1,S=0,I=1/2,J=0,1,2$ cases studied in \cite{raquel2} and $C=0,S=0,I=0,1,J=0,1,2$ studied in \cite{xyz}.
 
In order to calculate the $t$($u$)-channel vector meson exchange diagrams, one must project the amplitudes in s wave. This can be done by means of the following replacements:
\begin{eqnarray}
k_1\cdot k_2 &=& \frac{s-M^2_1-M^2_2}{2}\nonumber\\
k_1\cdot k_3 &=&k^0_1 k^0_3-\vec{p}\cdot \vec{q}\to \frac{(s+M^2_1-M^2_2) (s+M^2_3-M^2_4)}{4 s}\nonumber
\end{eqnarray}
where '$\to$' denotes the projection over s wave, and $k_1=(k^0_1,\vec{p})$, $k_2=(k^0_2,-\vec{p})$,  $k_3=(k^0_3,\vec{q})$, $k_4=(k^0_4,-\vec{q})$ and $M_i$, with $i=1,4$, is the mass of each external particle. 

After projecting the amplitudes in isospin, spin and s wave, they will be inserted into the Bethe-Salpeter equation as kernel $V$, which in the on-shell formalism \cite{Oller1,OsetRa} can be expressed  by
\begin{equation}\label{Bethe}
T=(\hat{1}-VG)^{-1}\,V\ .
\end{equation}
The kernel $V$ is of matrix type where its elements are the ($V_1(k_1)V_2(k_2)\to V_3(k_3)V_4(k_4)$) amplitudes in lowest order in $g^2$ evaluated above in the base of spin and isospin.
In Eq. (\ref{Bethe}), $G$ is a diagonal matrix with the two meson loop 
functions $G_i$ for each $V_1V_2$ channel:
\begin{equation}
G_i=i\int \frac{d^4 q}{(2\pi)^4}\frac{1}{q^2-M_1^2+i\eps}\frac{1}{(P-q)^2-M_2^2+i\eps}\ ,
\label{loop}
\end{equation}
which upon using dimensional regularization can be written as
\begin{eqnarray}
G_i&=&{1 \over 16\pi ^2}\biggr( \alpha +Log{M_1^2 \over \mu ^2}+{M_2^2-M_1^2+s\over 2s}
  Log{M_2^2 \over M_1^2}\nonumber\\ 
  &+&{p\over \sqrt{s}}\Big( Log{s-M_2^2+M_1^2+2p\sqrt{s} \over -s+M_2^2-M_1^2+
  2p\sqrt{s}}+Log{s+M_2^2-M_1^2+2p\sqrt{s} \over -s-M_2^2+M_1^2+  2p\sqrt{s}}\Big)\biggr)\ ,
  \label{dimreg}
\end{eqnarray}
 where $P$ is the total four-momentum of the two mesons and $p$ is the three-momentum 
 of the mesons in the center-of-mass frame:
 \begin{equation}
 p=\frac{\sqrt{(s-(M_1+M_2)^2)\,(s-(M_1-M_2)^2)}}{2\,\sqrt{s}}\ .
 \end{equation}
 Analogously, one can calculate the loop function by using a cut 
 off
 \begin{equation}
G_i=\int_0^{q_{max}} \frac{q^2 dq}{(2\pi)^2} \frac{\omega_1+\omega_2}{\omega_1\omega_2 [{(P^0)}^2-(\omega_1+\omega_2)^2+i\epsilon]   } \ ,\label{loopcut}
\end{equation}
 where $q_{max}$ stands for the cut off in the three-momentum, $\omega_i=(\vec{q}\,^2_i+M_i^2)^{1/2}$ 
 and the square of center-of-mass energy ${(P^0)}^2=s$. In the complex plane and  for a general $\sqrt{s}$, the loop function in the second Riemann sheet can be written as \cite{Roca}:
 \begin{equation}
G^{II}_i(\sqrt{s})=G^{I}_i(\sqrt{s})+i \frac{p}{4\pi \sqrt{s}}\,\hspace{1cm}Im(p)>0
\label{secondR}
\end{equation}
where $G^{II}_i$ refers to the loop function on the second Riemann sheet and $G^{I}_i$ is the loop function in the first Riemann sheet given by Eqs. (\ref{dimreg}) and (\ref{loopcut}) for each channel $i$.
%When looking for poles, 
Bound states appear as poles over the real axis and below thresholds on the first Riemann sheet. In contrast resonances are identified by poles on the second Riemann sheet above the thresholds of the channels which are open.

The channels that we consider are:
\begin{itemize}
\item $\mathbf{C=0;S=1;I=1/2}$ (hidden charm):
\begin{center}
$D_s^*\bar{D}^*(4121)$, $J/\psi K^*(3990)$
\end{center}
\item $\mathbf{C=1;S=-1;I=0}$ and $\mathbf{1}$:
\begin{center}
$D^*\bar{K}^*(2902)$
\end{center}
\item $\mathbf{C=1;S=1;I=0}$:
\begin{center}
$D^*K^*(2902)$, $D^*_s\omega(2895)$, $D^*_s\phi(3132)$
\end{center}
\item $\mathbf{C=1;S=1;I=1}$:
\begin{center}
$D^*K^*(2902)$, $D^*_s\rho(2888)$
\end{center}
\item $\mathbf{C=1;S=2;I=1/2}$:
\begin{center}
$D^*_s K^*(3006)$
\end{center}
\item $\mathbf{C=2;S=0;I=0}$ and $\mathbf{1}$:
\begin{center}
$D^* D^{*}(4017)$
\end{center}
\item $\mathbf{C=2;S=1;I=1/2}$:
\begin{center}
$D^*_s D^{*}(4121)$
\end{center}
\item $\mathbf{C=2;S=2;I=0}$:
\begin{center}
$D^*_s D^*_s(4224)$
\end{center}
\end{itemize}
 Here the quantities in parenthesis correspond to the sum of the masses of the two vector mesons.
\subsection{Convolution due to the vector meson mass distribution}

 In the channels $i=V_1V_2$, where  the width of one of the vector mesons involved is quite large, the mass distribution of the vector meson has to be taken into account. We demonstrate our technique by means of a broad $V_1$ meson. Its width is taken into account replacing the loop function $G$ in Eq. (\ref{Bethe}) by the convoluted $\tilde{G}$ \cite{hidekoroca}:
%Take for example $V_1$, the vector meson in a particular channel with a large width, then, this is taken into account in the function loop by  means of replacing $G$ in Eq. (\ref{Bethe}) by $\tilde{G}$ \cite{hidekoroca}:
\begin{eqnarray}
\tilde{G}(s)&=& \frac{1}{N}\int^{(M_1+2\Gamma_1)^2}_{(M_1-2\Gamma_1)^2}d\tilde{m}^2_1(-\frac{1}{\pi}) {\cal I}m\frac{1}{\tilde{m}^2_1-M^2_1+i\Gamma(\tilde{m})\tilde{m}_1} G(s,\tilde{m}^2_1,M_2^2)\ ,
\label{Gconvolution}
\end{eqnarray}
with
\begin{equation}
N=\int^{(M_1+2\Gamma_1)^2}_{(M_1-2\Gamma_1)^2}d\tilde{m}^2_1(-\frac{1}{\pi}){\cal I}m\frac{1}{\tilde{m}^2_1-M^2_1+i\Gamma(\tilde{m})\tilde{m}_1}\ ,
\label{Norm}
\end{equation}
where $M_1$ and $\Gamma_1$ are the nominal mass and width of the vector meson. $\Gamma(\tilde{m})$ is given by
\begin{equation}
\tilde{\Gamma}(\tilde{m})=\Gamma_0\frac{q^3_\mathrm{off}}{q^3_\mathrm{on}}\Theta(\tilde{m}-m_1-m_2)
\end{equation}
with
\begin{equation}\label{eq:mom}
q_\mathrm{off}=\frac{\lambda^{1/2}(\tilde{m}^2,m_1^2,m_2^2)}{2\tilde{m}},\quad
q_\mathrm{on}=\frac{\lambda^{1/2}(M_1^2,m_1^2,m_2^2)}{2 M_1}\,.
\end{equation}
In Eq. (\ref{eq:mom}), $m_1$, $m_2$ are the masses of the two pseudoscalar mesons in the decay $V_1(M_1)\to p_1(m_1)p_2(m_2)$. We only use Eq. (\ref{Gconvolution}) for the cases where a $\rho$ or $K^*(\bar{K}^*)$ meson are involved in a particular channel $i$. For the $\rho$ meson, $\Gamma_1=146.2$ MeV, and $m_1=m_2=m_\pi$ while for the $K^*$ meson we have
$\Gamma_1=50.55$ MeV and $m_1=m_K$, $m_2=m_\pi$.

The use of $\tilde{G}$ in Eq.~(\ref{Bethe}) provides larger widths of the states 
than using only $G$ (Eq. (\ref{loop})).

\subsection{Box diagrams}
The box diagrams containing intermediate states of two pseudoscalar mesons provide a mechanism to consider the two pseudoscalar decay mode of the dynamically generated resonances. In fact, these box diagrams were considered in \cite{raquel1,raquel2,geng,xyz}. The real part was negligible compared to the strong interaction obtained by means of the four-vector contact term plus vector-exchange diagrams of Fig. \ref{fig:fig1} a) and c). However, the imaginary part of the box diagrams is relevant for the generation of the width of the resonances. We will come to this issue later on.
\begin{figure}
\begin{center}
\includegraphics[width=15cm]{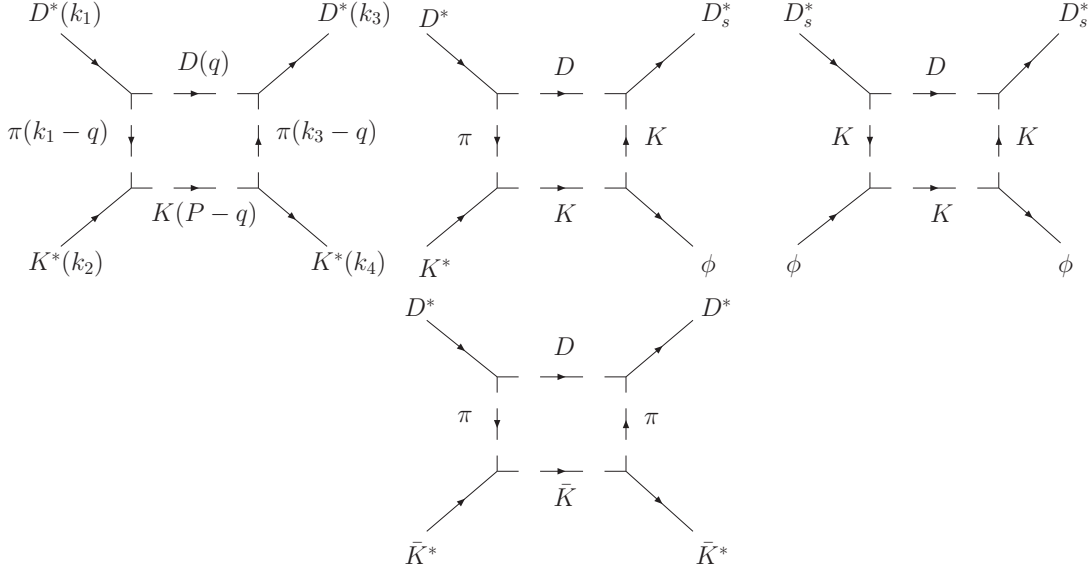}
\end{center}
\caption{Box diagrams included in the calculus.}
\label{fig:fig2} 
\end{figure}

\begin{figure}
\begin{center}
\includegraphics[width=5cm]{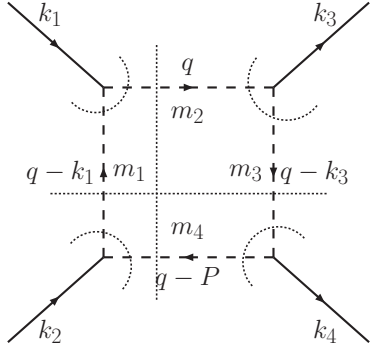}
\end{center}
\caption{Box diagram containing four pseudoscalar mesons. The cuts in the diagram provide the sources of imaginary part of the potential.}
\label{fig:box} 
\end{figure}

In Fig. \ref{fig:box} we represent the box diagram and its momentum variables. The vertices are provided by the same hidden gauge formalism (HGS) used in Section 2 by means of the Lagrangian
\begin{equation}
{\cal L}_{V\Phi\Phi}=-ig\langle V^\mu[\Phi,\partial_\mu \Phi]\rangle\ .
\label{lVPP}
\end{equation}
The generic structure of the diagram in Fig. \ref{fig:box} is:

\begin{eqnarray}
 V&\sim&C \int\frac{d^4q}{(2\pi)^4} \epsilon_1\cdot(2q-k_1) \epsilon_2\cdot (2q-k_3)\\
 && \times \epsilon_3\cdot(2q-k_3-P)\epsilon_4\cdot(2q-k_1-P)\nonumber\\
 &&\times\frac{1}{(q-k_1)^2-m_1^2+i\epsilon}
 \frac{1}{q^2-m_2^2+i\epsilon}\nonumber\\
 &&\times\frac{1}{(q-k_3)^2-m_3^2+i\epsilon}\frac{1}{(q-P)^2-m_4^2+i\epsilon},\nonumber
 \end{eqnarray}
 where $C$ is the coupling of a certain transition.
 The approximation of neglecting the three-momenta of the external particles leads to a simplified expression for $V$
  \begin{eqnarray}
 V&\sim&C_1 \int\frac{d^4q}{(2\pi)^4} \epsilon_1^i\epsilon_2^j\epsilon_3^m\epsilon_4^n q^i q^j q^m q^n\nonumber\\
 &&\times\frac{1}{(q-k_1^0)^2-m_1^2+i\epsilon}
 \frac{1}{q^2-m_2^2+i\epsilon}\nonumber\\
 &&\times\frac{1}{(q-k_3^0)^2-m_3^2+i\epsilon}\frac{1}{(q-P^0)^2-m_4^2+i\epsilon}\nonumber\\
 &=&C_1 G,
 \end{eqnarray}
 with $C_1=16C$.
This integral is logarithmically divergent and we regularize it with a cut off in the three momenta of natural size. Thus, the integral in $q^0$ is performed by means of the residue theorem and then the integral in the three momenta is calculated with a cut off of $q_{\mathrm{max}}=1.2$ GeV \cite{raquel1,raquel2,xyz}. We include these diagrams in the sectors where the interaction is strong enough to obtain bound states or resonances. Looking at the Tables in the Appendix, these sectors (and the channels involved) are:
\begin{itemize}
\item $C=1;S=-1;I=0;J=0,1$ and $2$:
\begin{center}
$D^*\bar{K}^*$
\end{center}
\item $C=1;S=1;I=0;J=0,1$ and $2$:
\begin{center}
$D^*K^*$, $D^*_s\phi$, $D^*_s\omega$
\end{center}
\item $C=1;S=1;I=1;J=0,1$ and $2$: 
\begin{center}
$D^*K^*$, $D^*_s\rho$
\end{center}
\item $C=2;S=0;I=0;J=1$:
\begin{center}
$D^*D^*$
\end{center}
\item $C=2;S=1;I=1/2;J=1$: 
\begin{center}
$D^*_s D^*$
\end{center}
\end{itemize}
However, the box diagrams only have a contribution for the quantum numbers $J^P=0^+$ and $2^+$. The reason is the following: the VV system has positive parity in s wave, which forces the PP intermediate state to be in $L=0,2$.  Since the two pseudoscalar meson do not have a spin, the only possibilities are $J^P=0^+$ and $2^+$. Hence we do not consider it for the last two sectors where $J=1$. For the other quantum numbers we consider the box diagrams in Fig. \ref{fig:fig2}. We do not include any box diagram for the channel $D_s^*\rho$ since $\rho$ goes to $\pi\pi$ and the vertex $D^*_s\pi D_s$ is equal to zero. Of course there exist other box diagrams involving the exchange in the t-channel of two pseudoscalars diferent from $\pi\pi$, $\pi K$ or $KK$ (the latter illustrated in Fig. \ref{fig:fig2}) but they are suppressed and can therefore be neglected. Crossed box diagrams (with four pseudoscalar mesons in the intermediate state) and box diagrams involving anomalous couplings 
 were also calculated in \cite{raquel1}, but they were found to be much smaller, especially in the case of the anomalous coupling, than the contributions from the box diagram of Fig. \ref{fig:box}. The final formula for each of the diagrams in Fig. \ref{fig:fig2} is given in the Appendix. One can see in these formulas that the cuts plotted in the diagram in Fig. \ref{fig:box} are clearly visible in the denominators.

Following the ideas of \cite{raquel2} we include two different form factors in the integral of the box-diagram potential (formulas of the Appendix). These are:
\begin{itemize}
\item Model A:
We multiply the vertices in the diagram of Fig. \ref{fig:box} by:
\begin{equation}\label{eq:form1}
F_1(q^2)=\frac{\Lambda_b^2-m_1^2}{\Lambda_b^2-(k_1^0-q^0)^2+|\vec{q}|^2},
\end{equation}
\begin{equation}\label{eq:form2}
F_3(q^2)=\frac{\Lambda_b^2-m_3^2}{\Lambda_b^2-(k_3^0-q^0)^2+|\vec{q}|^2},
\end{equation}
with $q^0=\frac{s+m_2^2-m_4^2}{2\sqrt{s}}$, $\vec{q}$ being the running variable,
and $\Lambda_b=1.4,1.5$ GeV~\cite{raquel1}.  These form factors
were inspired by the empirical form factors used in the decay of vector mesons~\cite{Titov:2000bn,Titov:2001yw}. Therefore, we add $F_1(q^2)^2 F_3(q^2)^2$ to the integrand in Eqs. (\ref{eq:boxdk}), (\ref{eq:boxdkdsphi}) and (\ref{eq:boxdsphi}) and we put $g=M_\rho/2\,f_\pi$. 
\item Model B:
Here we use a exponential parametrization for a off-shell $\pi(K)$ evaluated using QCD sum rules \cite{Navarra},
\begin{eqnarray}
F(q^2)=e^{((q^{0})^2-|\vec{q}|^2)/\Lambda^2}\ ,
\label{eq:formfactorprim}
\end{eqnarray}
with $\Lambda=1,1.2$ GeV and $q^0=\frac{s+m_2^2-m_4^2}{2\sqrt{s}}$. So we add $F(q^2)^4$ to the integrand in Eqs. (\ref{eq:boxdk}), (\ref{eq:boxdkdsphi}) and (\ref{eq:boxdsphi}). In this case we also change the factor $g^4$ in these equations by the corresponding product of g's, $g=M_\rho/2\,f_\pi$, with $f_\pi=93$ MeV, $g_{D_s}=M_{D^*_s}/2\,f_{D_s}=5.47$ with $f_{D_s}=273/\sqrt{2}$ MeV \cite{pdg} and $g_D=g^{\mathrm{exp}}_{D^*D\pi}=8.95$ (experimental value) \cite{Ahmed,Belyaev,Colangelo}.
\end{itemize}

\begin{figure}
\begin{center}
\includegraphics[width=16cm]{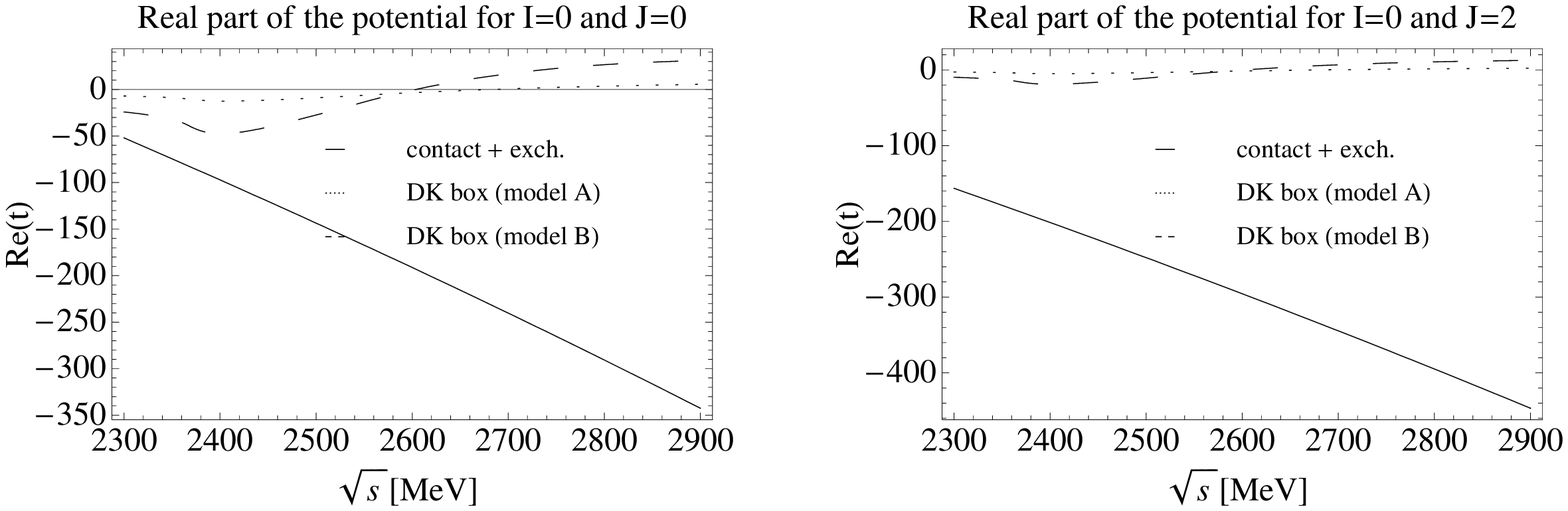}
\end{center}
\caption{Comparison of the real part of the box diagram with the contact term plus vector-exchange term for the $D^*K^*\to D^*K^*$ amplitude and $I=0$, $J=0$ and $J=2$ respectively.}
\label{fig:realbox} 
\end{figure}
\begin{figure}
\begin{center}
\includegraphics[width=16cm]{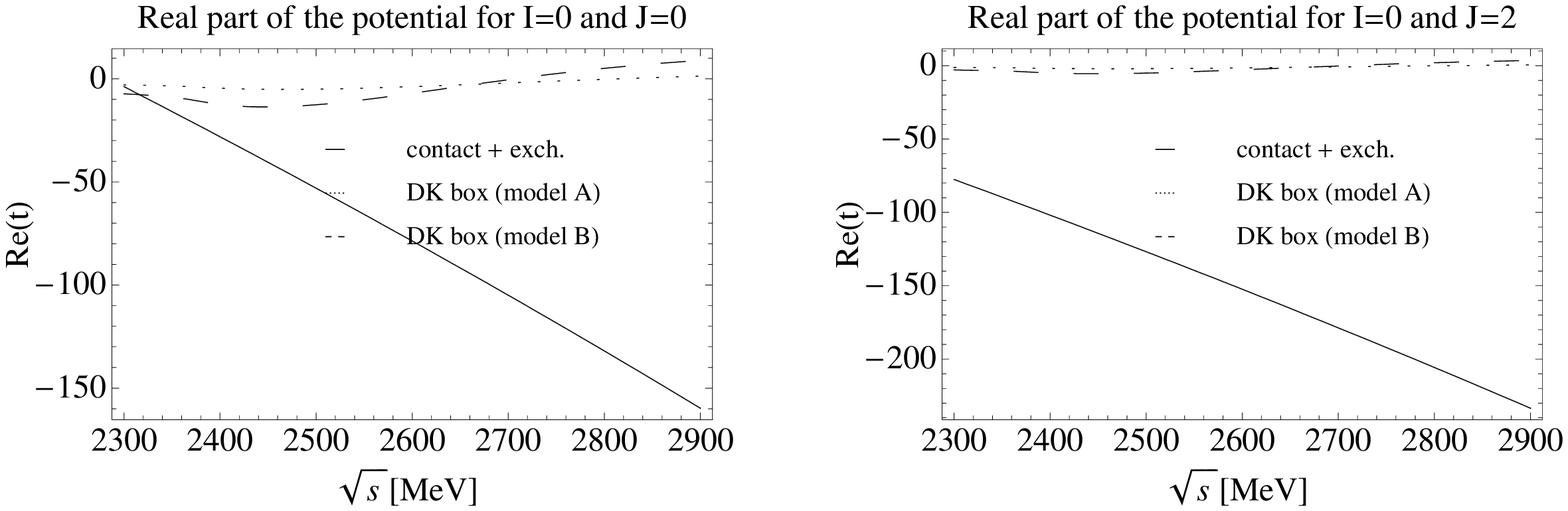}
\end{center}
\caption{Comparison of the real part of the box diagram with the contact term plus vector-exchange term for the $D^*K^*\to D_s^*\phi$ amplitude and $I=0$, $J=0$ and $J=2$ respectively.}
\label{fig:realbox1} 
\end{figure}
\begin{figure}
\begin{center}
\includegraphics[width=16cm]{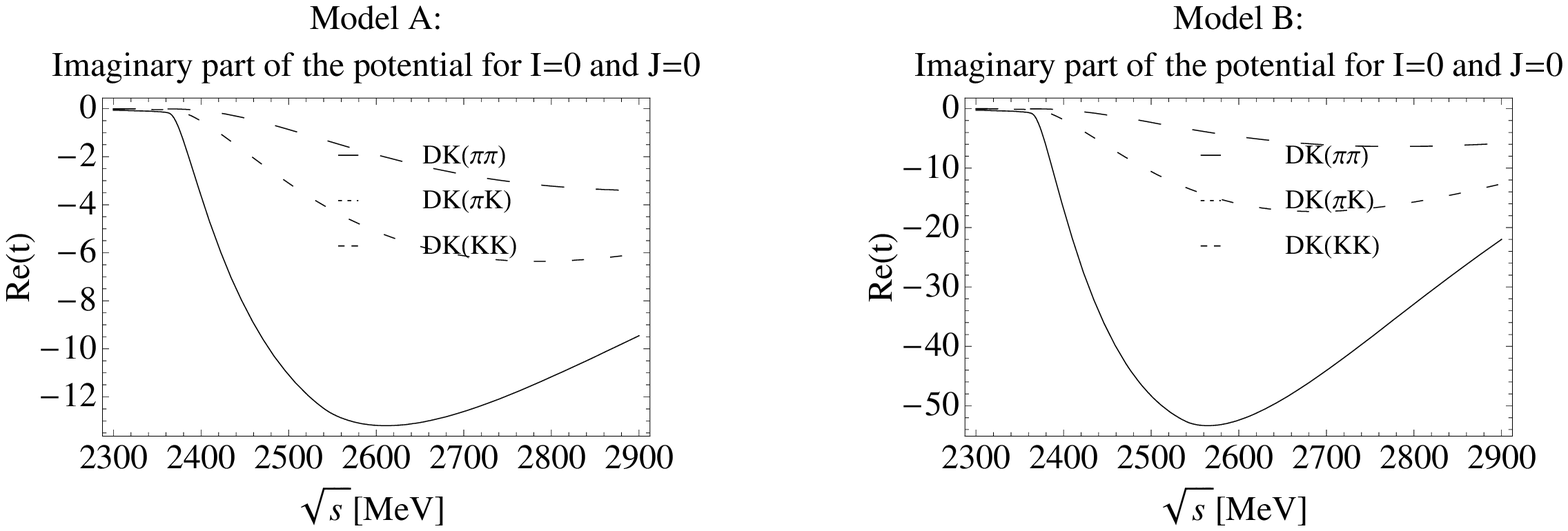}
\end{center}
\caption{Imaginary part of the box diagrams in Fig. \ref{fig:fig2} for $I=0$ and $J=0$.}
\label{fig:imagbox} 
\end{figure}

In Figs. \ref{fig:realbox} and \ref{fig:realbox1} we compare the real parts of the  box diagrams with the contact terms plus vector-exchange terms for the $D^*K^*\to D^*K^*$ and $D^*K^*\to D^*_s\phi$ amplitudes (the interaction is very attractive for these amplitudes, see Table \ref{tab:p3}). As one can see in this figure, the box diagram has a small real part compared to the strong potential provided by the four-vector contact terms plus vector-exchange diagrams, particularly in the region of energies corresponding to the states that we find. Therefore, one can neglect the real part of the box diagrams as it was done in \cite{raquel1,raquel2,xyz}. In Fig. \ref{fig:imagbox} we depict the imaginary part of the box diagrams in Fig. \ref{fig:fig2} for the two models. Here we set $\Lambda=1400$ MeV for the Model A, while we put $\Lambda=1200$ MeV when using Model B. As this figure shows, the Model B with the form factor of Eq. (\ref{eq:formfactorprim}) provides a larger imaginary
  part compared to Model A which results in a larger width of the resonance.

\section{Results}
In this section we will present the results for each sector as follows: First, we apply the Bethe-Salpeter equation Eq. (\ref{Bethe}), by taking the potential $V$ from the Tables in the Appendix (contact terms plus vector-exchange diagrams). Here, we use the following parameters: $g=M_\rho/2\,f_\pi$, we fix $\mu=1500$ MeV for all the sectors and set the subtraction constant $\alpha=-1.6$ (value very close the one used in \cite{Gamermann2}, $-1.55$, and \cite{raquel2}, $-1.74$) in the sectors $C=1;S=-1$, $C=1;S=1$ and $C=1;S=2$. Note that $\mu$ and $\alpha$ are not independent which justifies the determination of $\mu$ and then adjusting $\alpha$ to the data. In the other sectors, $C=0;S=1$ (hidden charm), $C=2;S=0$, $C=2;S=1$ and $C=2;S=2$, we put $\alpha=-1.4$. The reason is that we use a different set of the parameters $\mu$ and $\alpha_H$ in comparison to the earlier study of the dynamically generated $D^\ast_{(s)}\bar D^\ast_{(s)}$ resonances in \cite{xyz} with $\mu=1000$
  MeV and $\alpha_H=$-2.07. In the present approach we set $\mu=1500$ MeV as in \cite{Gamermann1,Gamermann2,raquel2} and have to adapt $\alpha_H$ accordingly in order to be able to reproduce the XYZ states in \cite{xyz}.
%This last value of $\alpha$ is chosen for the following reason: the sector with two charmed mesons was studied in \cite{xyz} . There the authors use other different parameters, $\mu=1000$ MeV to reproduce the results of \cite{geng} and $\alpha_H=-2.07$. Thus, in this work, where we take a different value of $\mu$, $\mu=1500$ MeV as in \cite{Gamermann1,Gamermann2,raquel2}, we choose $\alpha$ to reproduce the XYZ states obtained in \cite{xyz}. }
Then, we evaluate the pole positions in the sectors where we find attractive interaction and calculate the couplings to each channel from the residue of the amplitudes, since, close to a pole, the amplitudes from Eq. (\ref{Bethe}) look like
\begin{equation}
T_{ij}\approx \frac{g_i g_j}{s-s_{p}}\ .
\label{poleT}
\end{equation}
Therefore, the constants $g_i$ ($i=VV$ channel), which provide
the couplings of the resonance 
to the particular channels can be calculated by means of the residues of
the amplitudes. The pole positions and couplings are given in Tables \ref{tab:res0}, \ref{tab:res1}, \ref{tab:res2}, \ref{tab:res3} and \ref{tab:res4}. Then we replace the expression for $G$ of Eq. (\ref{dimreg}) by the convoluted $\tilde{G}$ of Eq. (\ref{Gconvolution}) and additionally include the box diagrams in Fig. \ref{fig:fig2}. These modifications do not practically change the positions of the poles and the couplings are barely affected. However, the convolution of the mass distribution and the consideration of the pseudoscalar decay channels in terms of box diagrams leads to a larger width of the respective resonances. The reevaluation of the Bethe-Salpeter equation, Eq. (\ref{Bethe}), leads to the squared transition amplitudes pictured in the Figs. \ref{fig:T00m1}, \ref{fig:T02m1}, \ref{fig:T00}, \ref{fig:T02} and \ref{fig:T12}. The corresponding masses and widths are given in Tables \ref{tab:wid0}, \ref{tab:wid1} and \ref{tab:wid2}.

%We then proceed to replace the include the convoluted $\tilde{G}$ in Eq. (\ref{Gconvolution}) instead of $G$, Eq. (\ref{dimreg}), and also the box diagrams in Fig. \ref{fig:fig2}, what does not change practically the positions of the poles, hence, the constants $g_i$ either, as discussed in Section 2.2, reevaluating Eq. (\ref{Bethe}) and plotting the squared amplitudes in the Figs. \ref{fig:T00}, \ref{fig:T02}, \ref{fig:T12}, \ref{fig:T00m1} and \ref{fig:T02m1}. Finally, real masses and widths from these plots are given for both models in Section 2.2, A and B. We discuss the results independently for each sector in what follows.

\subsection{$C=0;S=1;I=1/2$ (hidden charm)}

The amplitudes from the four-vector contact terms plus vector-exchange diagrams can be found in Table \ref{tab:p0} in the Appendix. We can see from the tables that the potential is small and repulsive except for the $D^*_s\bar{D}^*\to J/\psi K^*$ and $D^*_s\bar{D}^*\to D^*_s\bar{D}^*$ amplitudes for $J=1$ and $2$ respectively. However, the attraction is too small to bind the system and therefore we do not get poles or possible states from the T-matrix.

\subsection{$C=1;S=-1;I=0$}
\begin{figure}
\begin{center}
\includegraphics[width=16cm]{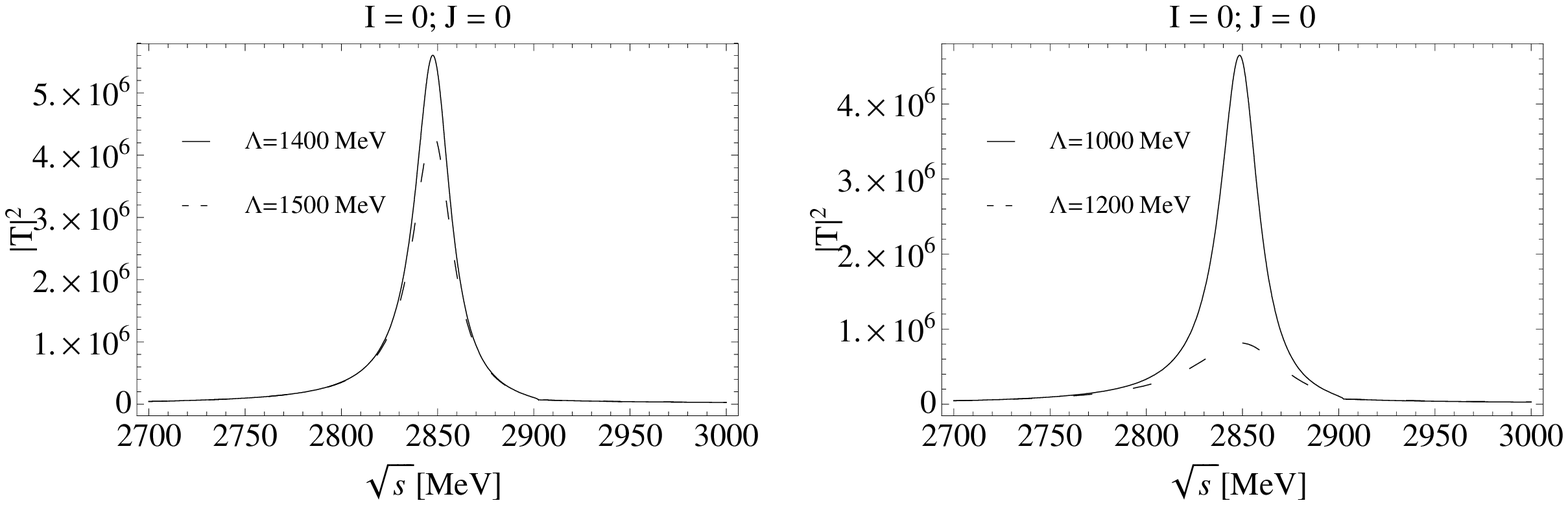}
\end{center}
\caption{Squared amplitude in the $D^*\bar{K}^*$ channel for $I=0$ and $J=0$. Left: Model A, right: Model B.}
\label{fig:T00m1} 
\end{figure}

\begin{figure}
\begin{center}
\includegraphics[width=16cm]{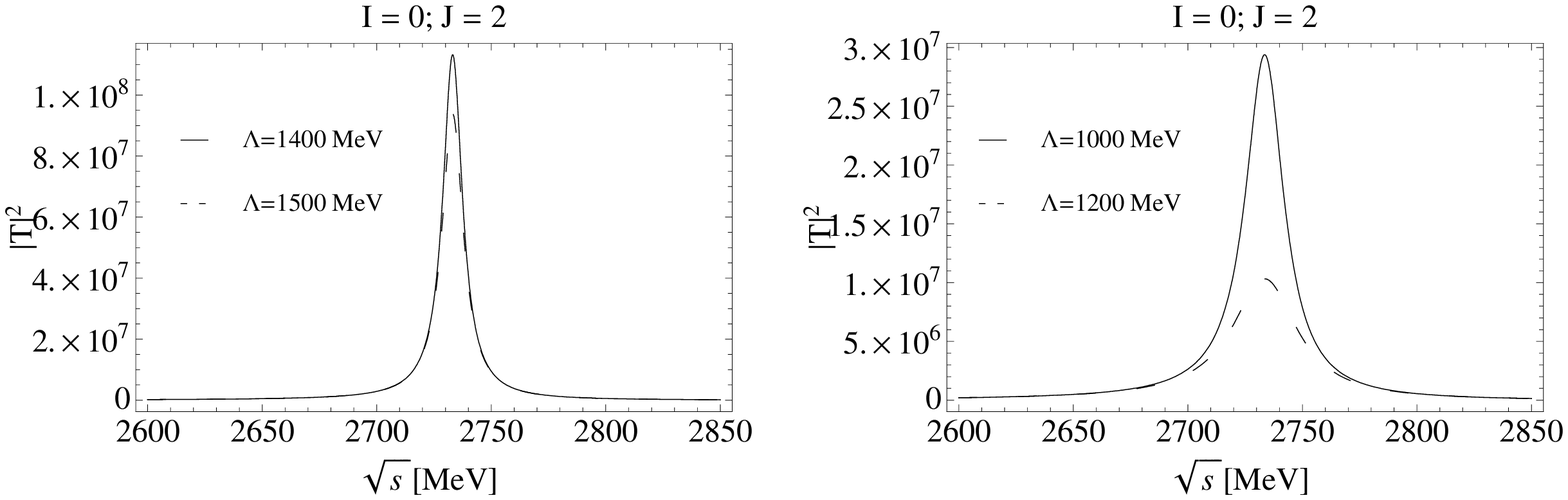}
\end{center}
\caption{Squared amplitude in the $D^*\bar{K}^*$ channel for $I=0$ and $J=2$. Left: Model A, right: Model B.}
\label{fig:T02m1} 
\end{figure}

 In contrast to the above sector the potential in the case of $C=1$ and $S=-1$ is very attractive as indicated in Table \ref{tab:p1}. For $I=0$ and $J=0,1$ the potential is around $-10\, g^2$ whereas it is about $-16\,g^2$ for $J=2$.
In this sector the strong interaction from the potential leads to bound states. We obtain one resonance for each spin, $J=0,1$ and $2$, where the corresponding pole positions and couplings are given in Table \ref{tab:res0}. The convolution of the $G$ function due to the $\bar{K}^*$ width leads to a minor shift in the pole positions (only $3$ MeV for $J=2$) and around $3$ MeV in the widths for the three states. This is a minor effect compared to the contribution of the box diagrams. Therefore we neglect the $K^\ast$ width in the final $|T|^2$ analysis. $|T|^2$ is depicted in Figs. \ref{fig:T00m1} and \ref{fig:T02m1} for $J=0$ and $2$ and for the two models A and B after the inclusion of the corresponding box diagrams of Fig. \ref{fig:fig2}. Here, the two models lead to similar results except for the model B with $\Lambda=1200$ MeV. In Table \ref{tab:wid0} we show the values of the masses and final widths of the states. Since these states have exotic flavor quantum numbers ther
 e is no possible $q\bar{q}$ counterpart.
\begin{table}[htpb]
 \begin{center}
\begin{tabular}{cc|c}
$I[J^{P}]$& $\sqrt{s}_{\mathrm{pole}}$ (MeV)\T\B&$g_{D^*\bar{K}^*)}$\\
\hline
\hline
 $0[0^{+}]$&$2848$\T\B&$12227$\\
 $0[1^{+}]$&$2839$\T\B&$13184$\\
 $0[2^{+}]$&$2733$\T\B&$17379$\\
\hline
\end{tabular}
\end{center}
\caption{$C=1;S=-1;I=0$. Quantum numbers, pole positions and couplings $g_{i}$ in units of MeV. Here, $\alpha=-1.6$.}
\label{tab:res0}
\end{table} 

\begin{table}[htpb]
 \begin{center}
\begin{tabular}{cc|cc}
$I[J^{P}]$& $\sqrt{s}_{\mathrm{pole}}$ (MeV)\T\B& Model & $\Gamma$ (MeV)\\
\hline
\hline
$0[0^{+}]$&$2848$\T\B& A, $\Lambda=1400$ MeV&$23$\\
&\T\B& A, $\Lambda=1500$ MeV&$30$\\
&\T\B& B, $\Lambda=1000$ MeV&$25$\\
&\T\B& B, $\Lambda=1200$ MeV&$59$\\
\hline
$0[1^{+}]$&$2839$\T\B& Convolution &$3$\\
\hline
$0[2^{+}]$&$2733$\T\B& A, $\Lambda=1400$ MeV&$11$\\
&\T\B& A, $\Lambda=1500$ MeV&$14$\\
&\T\B& B, $\Lambda=1000$ MeV&$22$\\
&\T\B& B, $\Lambda=1200$ MeV&$36$\\
\hline
\end{tabular}
\end{center}
\caption{$C=1;S=-1;I=0$. Mass and width for the states with $J=0$ and $2$.}
\label{tab:wid0}
\end{table} 

\subsection{$C=1;S=-1;I=1$}

In this sector, which also has exotic quantum numbers, we can see from Table \ref{tab:p2} that the interaction is very repulsive in contrast to the previous case of $I=0$. Therefore, no bound states or resonances are found in this sector.
 
\subsection{$C=1;S=1;I=0$}
 
 The strong interaction coming from the contact terms plus vector-exchange diagrams leads to a potential of the order of $-18\,g^2$ to $-26\,g^2$, see Table \ref{tab:p3} in the Appendix, which is enough to bind the $D^*$ and $K^*$ mesons. In this sector we obtain three poles with masses $M=2683$, $2707$ and $2572$ MeV for $J=0$, $1$ and $2$, respectively. The potentials in Tab. \ref{tab:p3} provide the kernel $V$ of Eq. (\ref{Bethe}) which results in the pole positions and couplings summarized in Table \ref{tab:res1}. The state with $J=2$ is more bound than the other poles for $J=0$ and $1$ which can be identified with the $D^*_2(2573)$ resonance in the PDG. Here, the $D^*K^*$ channel is dominant for the three different spins. Nevertheless the other channels, $D^*_s\omega$ and $D^*_s\phi$ are not negligible.

\begin{figure}
\begin{center}
\includegraphics[width=16cm]{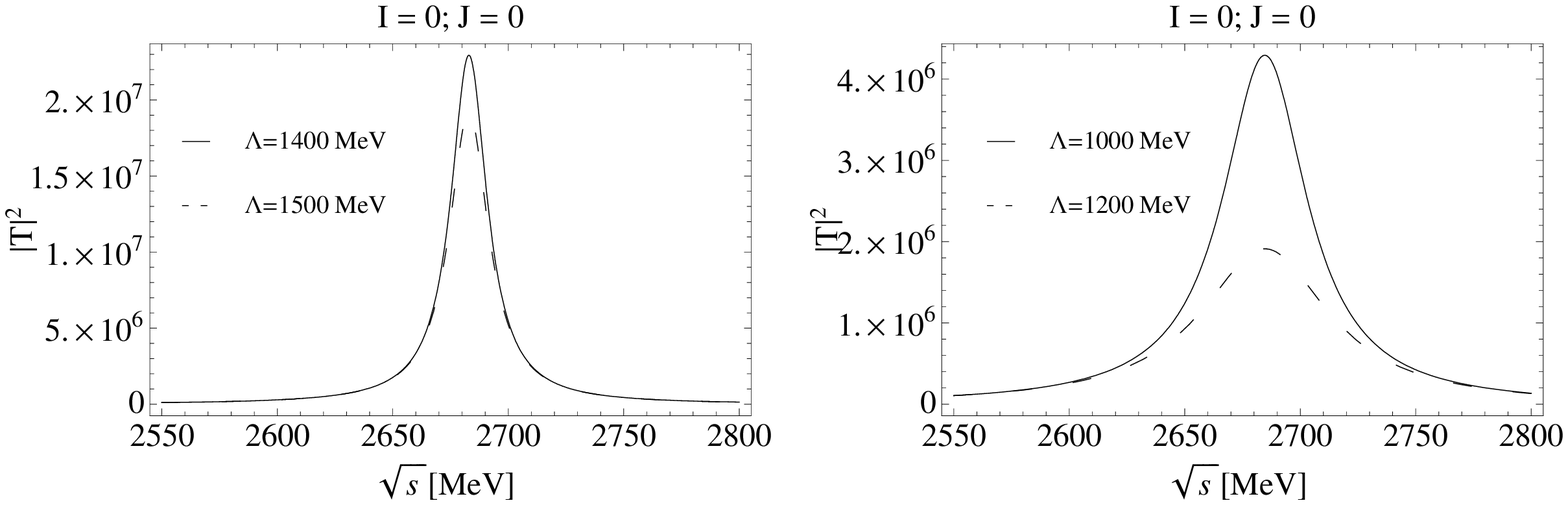}
\end{center}
\caption{Squared amplitude in the $D^*K^*$ channel for $I=0$ and $J=0$. Left: Model A, right: Model B.}
\label{fig:T00} 
\end{figure}

\begin{figure}
\begin{center}
\includegraphics[width=16cm]{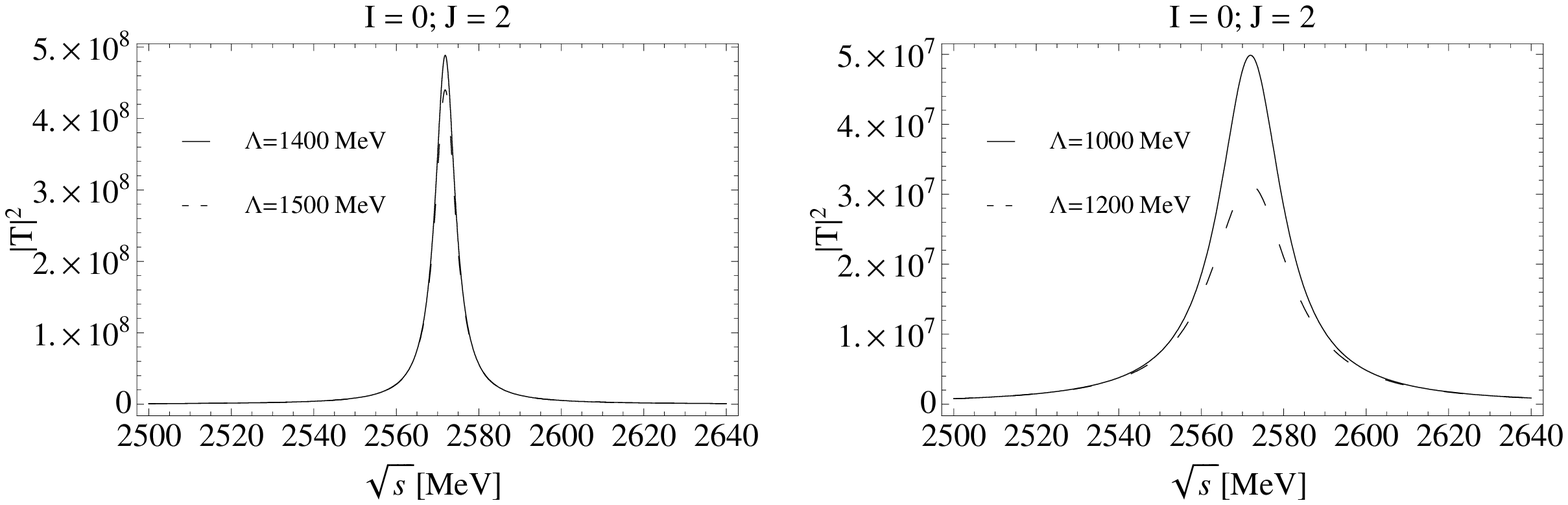}
\end{center}
\caption{Squared amplitude in the $D^*K^*$ channel for $I=0$ and $J=2$. Left: Model A, right: Model B.}
\label{fig:T02} 
\end{figure}

When considering the $K^*$ width, which is equivalent to replacing  $G$ by the convoluted $\tilde{G}$, neither the mass changes significantly (in fact only $2$ MeV) nor the width is affected by this modification. Therefore, the effect of the convolution is so small that it does not need to be considered. Only the consideration of the box diagrams has some influence on the  width. In Figs. \ref{fig:T00} and \ref{fig:T02} $|T|^2$ is plotted after the inclusion of the box diagrams of Fig. \ref{fig:fig2} for the two models A and B. We observe that these diagrams provide some width for the states with $J=0$ and $2$ (possible quantum numbers of the box diagrams), although the width provided by the model B is much bigger than that from model A. The values of the masses and widths are given in Table \ref{tab:wid1}. Model B with $\Lambda=1000$, $1200$ MeV provides a width for the state appearing around $2572$ MeV of $18-23$ MeV. 

We associate this state with the $D^*_{s2}(2573)$ of the PDG \cite{pdg} since the quantum numbers, position and width agree with those of the PDG. We should note that this is the case where we found the largest attraction, of the order of $-26\,g^2$, which is even bigger than what was found for $I=0,J=2$ in the $\rho\rho$ interaction ($\simeq -20\,g^2$) which lead to the production of the $f_2(1270)$ \cite{raquel1,geng}.
\begin{table}[htpb]
 \begin{center}
\begin{tabular}{cc|ccc}
$I[J^{P}]$& $\sqrt{s}_{\mathrm{pole}}$ (MeV)\T\B&$g_{D^*K^*}$&$g_{D^*_s\omega}$&$g_{D^*_s\phi}$\\
\hline
\hline
$0[0^+]$&$2683$\T\B&$15635$&$-4035$&$6074$\\
$0[1^+]$&$2707$\T\B&$14902$&$-5047$&$4788$\\
$0[2^+]$&$2572$\T\B&$18252$&$-7597$&$7257$\\
\hline
\end{tabular}
\end{center}
\caption{$C=1;S=1;I=0$. Quantum numbers, pole positions and couplings $g_{i}$ in units of MeV for $I=0$. Here $\alpha=-1.6$.}
\label{tab:res1}
\end{table} 
\begin{table}[htpb]
 \begin{center}
\begin{tabular}{cc|ccc}
$I[J^{P}]$& $\sqrt{s}_{\mathrm{pole}}$ (MeV)\T\B& Model & $\Gamma_{\mathrm{theo}}$ (MeV)& $\Gamma_{\mathrm{exp}}$ (MeV)\\
\hline
\hline
$0[0^+]$&$2683$\T\B& A, $\Lambda=1400$ MeV&$20$&-\\
&\T\B& A, $\Lambda=1500$ MeV&$25$&\\
&\T\B& B, $\Lambda=1000$ MeV&$44$&\\
&\T\B& B, $\Lambda=1200$ MeV&$71$&\\
\hline
$0[1^+]$&$2707$& Convolution &$4\times 10^{-3}$&-\T\B\\
\hline
$0[2^+]$&$2572$\T\B& A, $\Lambda=1400$ MeV&$7$&$\mathbf{20\pm 5}$ \cite{pdg}\\
&\T\B& A, $\Lambda=1500$ MeV&$8$&\\
&\T\B& \textbf{B}, $\mathbf{\Lambda=1000}$ \textbf{MeV}&$\mathbf{18}$&\\
&\T\B& \textbf{B}, $\mathbf{\Lambda=1200}$ \textbf{MeV}&$\mathbf{23}$&\\
\hline
\end{tabular}
\end{center}
\caption{$C=1;S=1;I=0$. Mass and width for the states with $J=0$ and $2$.}
\label{tab:wid1}
\end{table} 

\subsection{$C=1;S=1;I=1$}

\begin{figure}
\begin{center}
\includegraphics[width=16cm]{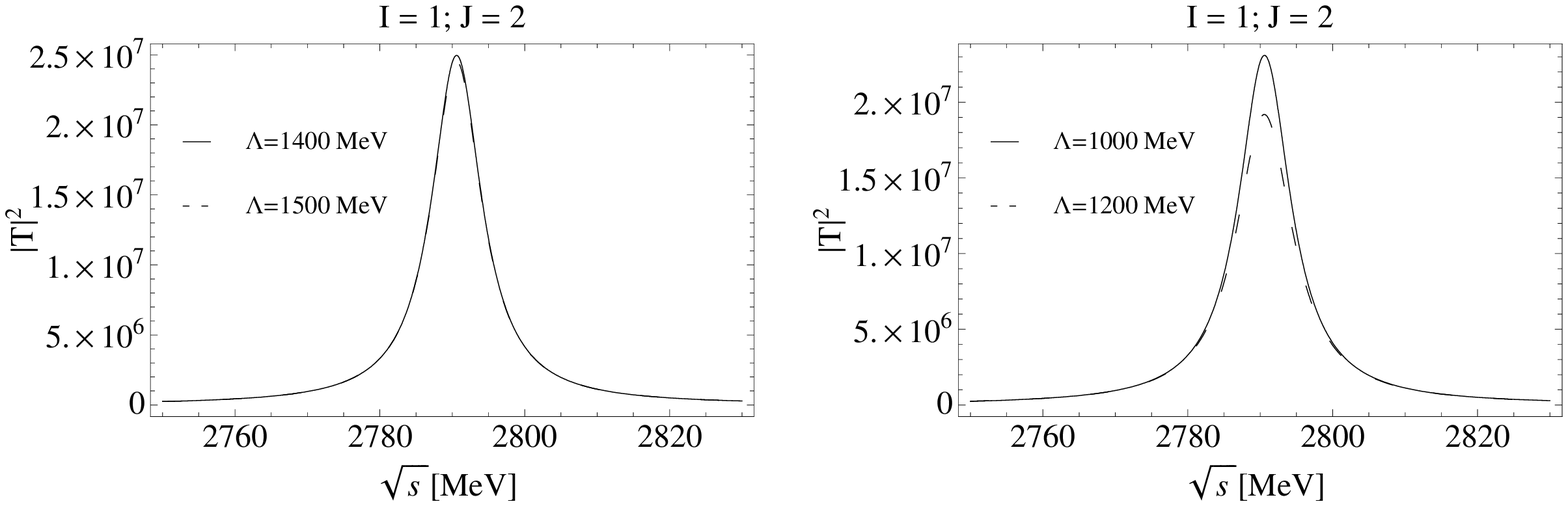}
\end{center}
\caption{Squared amplitude in the $D^*K^*$ channel for $I=1$ and $J=2$. Left: Model A, right: Model B.}
\label{fig:T12} 
\end{figure}

In this sector the potential is attractive for the $D^*K^*\to D^*_s\rho$ reaction. For $J=0$ and $1$ this potential is around $-7\,g^2$ whereas it is by a factor of two bigger $-13\,g^2$ for $J=2$ (see Table \ref{tab:p4}). In fact, we only obtain a pole for $J=2$. For $J=0$ and $1$ we only observe a cusp in the $D^*_s\rho$ threshold. In Table \ref{tab:res2} we show the pole position and couplings to the different channels. Both channels, $D^*K^*$ and $D^*_s\rho$, are equally important as one can deduce from the corresponding couplings. The broad width of the $\rho$ meson has to be taken into account by means of Eq. (\ref{Gconvolution}) which results in a width of $8$ MeV.  In this case the box diagrams in Fig. \ref{fig:fig2} for the $D^*K^*$ channel only make a small contribution to the width of the resonance (see Fig. \ref{fig:T12}). In contrast to the previous situations the width of the resonance is mainly generated by the convolution of the $\rho$ mass while the box diagr
 ams play a minor role. In Table \ref{tab:wid2} we give the exact values of the width in the two models which give very similar results. No experimental counterpart is found for this state in the PDG.
\begin{table}[htpb]
 \begin{center}
\begin{tabular}{cc|cc}
$I^G[J^{PC}]$& $\sqrt{s}_{\mathrm{pole}}$ (MeV)\T\B&$g_{D^*K^*}$&$g_{D^*_s\rho}$\\
\hline
\hline
 $1[2^+]$&$2786$\T\B&$11041$&$11092$\\
\hline
\end{tabular}
\end{center}
\caption{$C=1;S=1;I=1$. Quantum numbers, pole positions and couplings $g_{i}$ in units of MeV. Here $\alpha=-1.6$.}
\label{tab:res2}
\end{table} 

\begin{table}[htpb]
 \begin{center}
\begin{tabular}{cc|cc}
$I[J^{P}]$& $\sqrt{s}_{\mathrm{pole}}$ (MeV)\T\B& Model & $\Gamma$ (MeV)\\
\hline
\hline
$1[2^+]$&$2786$\T\B& A, $\Lambda=1400$ MeV&$8$\\
&\T\B& A, $\Lambda=1500$ MeV&$9$\\
&\T\B& B, $\Lambda=1000$ MeV&$9$\\
&\T\B& B, $\Lambda=1200$ MeV&$11$\\
\hline
\end{tabular}
\end{center}
\caption{$C=1;S=1;I=1$. Mass and width for the state with $J=1$ and $2$.}
\label{tab:wid2}
\end{table} 

\subsection{$C=1;S=2;I=1/2$}

This sector is exotic since a double-strange state is not reached in $q\bar{q}$. As we can see from Table \ref{tab:p5} in the Appendix, the potential is repulsive for all possible spins. Therefore we do not get any bound state or resonance in this sector. 

\subsection{$C=2;S=0;I=0$}

 In this case we study double charmed states by coupled $D^\ast D^\ast$ channels. The amplitudes are given in Table \ref{tab:p6}, where the potential is zero for $J=0$ and $2$. This can be explained by the fact that the $D^\ast D^\ast$ state is antisymmetric for $I=0$. Therefore, the only possibility to obtain a fully symmetric wave function is provided by $J=1$ which is equivalent to the rule $L+\tilde{S}+I=\mathrm{odd}$,  since $L=0$ for $s-$wave ($\tilde{S}$, spin $\equiv J$ for $L=0$). For $J=1$ the interaction is strongly attractive and we obtain a pole in the scattering matrix. The pole position and coupling to the $D^*D^*$ channel is given in Table \ref{tab:res3}. The width of the $D^*$ meson is very small ($\sim 100$ keV or less in the case of the neutral charmed meson), hence, we do not perform the convolution of the $G$ function. Since we deal with a $J=1$ state the inclusion of the box diagrams can be ruled out. Therefore we obtain a state with zero width or a very
  narrow width when considering the convolution. This sector with $C=2$ is exotic and so far there are no experimental observations.
\begin{table}[htpb]
 \begin{center}
\begin{tabular}{cc|c}
$I[J^{P}]$& $\sqrt{s}_{\mathrm{pole}}$ (MeV)\T\B&$g_{D^*D^*}$\\
\hline
\hline
 $0[1^+]$&$3969$\T\B&$16825$\\
\hline
\end{tabular}
\end{center}
\caption{$C=2;S=0;I=0$. Quantum numbers, pole positions and couplings $g_{i}$ in units of MeV. Here $\alpha=-1.4$.}
\label{tab:res3}
\end{table} 
\subsection{$C=2;S=0;I=1$}

Here we deal with the reversed situation as in the previous $I=0$ sector. The isospin combination for $I=1$ of the $D^*D^*$ channel is symmetric and therefore  $J=1$ is forbidden ($L+\tilde{S}+I=\mathrm{even}$). However, the potential is very repulsive for $J=0$ and $J=2$ (see Table \ref{tab:p7}) and consequently we do not obtain any pole in the scattering matrix.

\subsection{$C=2;S=1;I=1/2$}
This sector is also exotic. The amplitudes from the four-vector contact terms plus vector-exchange diagrams lead to a repulsive potential for $J=0$ and $2$ and is attractive for $J=1$ as indicated Table \ref{tab:p8}. We get a pole almost at the $D^*_sD^*$ threshold ($4121$ MeV), where the pole position and the coupling is given in Table \ref{tab:res4}. This state comes with zero width since the box diagrams are not possible for $J=1$ and any possible convolution of the $G$ function would lead to a very small width. This state is also a prediction of the model and needs to be confirmed by experiment. 
\begin{table}[htpb]
 \begin{center}
\begin{tabular}{cc|c}
$I[J^{P}]$& $\sqrt{s}_{\mathrm{pole}}$ (MeV)\T\B&$g_{D_s^*D^*}$\\
\hline
\hline
 $1/2[1^+]$&$4101$\T\B&$13429$\\
\hline
\end{tabular}
\end{center}
\caption{$C=2;S=1;I=1/2$. Quantum numbers, pole positions and couplings $g_{i}$ in units of MeV . Here, $\alpha=-1.4$.}
\label{tab:res4}
\end{table} 

\subsection{$C=2;S=2;I=0$}

 The $D^*_sD^*_s$ channel allows us to study double-charm double-strange objects. Since we deal with two identical particles with isospin zero, the isospin $D^*_sD^*_s$-state is symmetric and hence we get interaction for $J=0$ and $2$ while the potential zero for $J=1$ (see Table \ref{tab:p9}). Since the potential is strongly repulsive we do not obtain any state in this sector.

 In Table \ref{tab:summary} we give a summary of the states obtained together with the only experimental counterpart observed so far.

 \begin{table}[htpb]
 \begin{center}
 \begin{tabular}{lc|ccc|ccc}
$C,S$&$I[J^P]$&$\sqrt{s}$&$\Gamma_{\mathrm{A}}(\Lambda=1400)$&$\Gamma_{\mathrm{B}}(\Lambda=1000)$& State &$\sqrt{s}_{\mathrm{exp}}$&$\Gamma_{\mathrm{exp}}$\T\B\\
 \hline
 \hline
$1,-1$&$0[0^{+}]$&$2848$&$23$&$25$& & &\T\B\\
&$0[1^{+}]$&$2839$&$3$&$3$& & &\T\B\\
&$0[2^{+}]$&$2733$&$11$&$22$& & &\T\B\\
$1,1$&$0[0^+]$&$2683$&$20$&$44$& & &\T\B\\
&$0[1^+]$&$2707$&$4\times 10^{-3}$&$4\times 10^{-3}$& & &\T\B\\
&$0[2^+]$&$2572$&$7$&$18$&$D_{s2}(2573)$&$2572.6\pm0.9$&$20\pm5$\T\B\\
&$1[2^+]$&$2786$&$8$&$9$& & &\T\B\\
$2,0,$&$0[1^+]$&$3969$&$0$&$0$& & &\T\B\\
$2,1$&$1/2[1^+]$&$4101$&$0$&$0$& & &\T\B\\
\hline
 \end{tabular}
  \end{center}
  \caption{Summary of the nine states obtained. The width is given for the model A, $\Gamma_A$, and B, $\Gamma_B$. All the quantities here are in MeV.}
\label{tab:summary}
 \end{table}
\section{Conclusions}
 We studied dynamically generated resonances from vector-vector interaction in the charm-strange and hidden-charm sectors and extended for the first time the formalism to flavor exotic sectors. The hidden gauge Lagrangians provide a consistent method to include vector meson interaction in the coupled channel unitarity formalism. Our analysis of the $T$ matrix resulted in nine bound states. At the beginning these states appear with zero width (poles on the real axis). There are two effects which are relevant for the generation of the width of the resonance. First, the widths of the vector mesons involved need to be considered by the convolution of the two-meson loop function. This effect is  in particular important for the $D^*_s\rho$ channel. Second, the $PP$ decay modes of the vector mesons play an important role. In the present coupled channel approach   this issue is taken into account by the insertion of box diagrams with pseudoscalar mesons in the intermediate state. The
 se modifications lead to appreciable widths of the states which are close to the experimental observations if available. In the present work we can assign one resonance to an experimental counterpart, which is the $D^*_2(2573)$. For $C=1,S=1$ we obtain three states with masses $M=2683$, $2707$ and $2572$ MeV for $I=0$ and $J=0,1,2$ respectively. The widths lie around $44$, $0$ and $18$ MeV correspondingly (Model B with $\Lambda=1000$ MeV). We associate the state for $J=2$ with the $D^*_2(2573)$ giving a novel interpretation for this resonance as a vector - vector molecular state. The assumption of this structure is consistent with the $DK$ nature assumed for the $D^*(2317)$, the $D^*K$ molecular structure of the $D^*(2460)$ or the X(3872) ($D \bar{D}^\ast$). The other two states around $2700$ MeV are predictions of the model without experimental evidence for these masses and quantum numbers up to now. For $I=1$ we find only one state, of non exotic nature, a $2^+$ state arou
 nd $2786$ MeV.
 
 In the flavor-exotic sectors which have not been studied before, we obtain interesting predictions for new states. In the sector $C=1;S=-1;I=0$ we obtain three new exotic states with masses $M=2848$, $2839$ and $2733$ MeV and widths around $\Gamma=25$, $3$ and $22$ MeV, for the quantum numbers $I[J^{P}]=0[0^{+}]$, $0[1^{+}]$ and $0[2^{+}]$ respectively. In the case of the double-charm sectors $C=2;S=0;I=0$ and $C=2;S=1;I=1/2$ the potential leads to a bound system for $J=1$ only. That is, we deal with two very narrow states with masses around $M=3969$ and $4101$ MeV close to the thresholds of $D^*D^*$ and $D^*_sD^*$ respectively. In summary, all states are relatively narrow. For the quantum numbers $J^P=0^+,2^+$ the widths are lower than $71$ MeV (depending on the model) while all states with $J^P=1^+$ come with practically no width since the box diagrams do not contribute. There is no experimental counterpart for all exotic structures which can be considered as $D^*K^*$, $D^
 *\bar{K}^*$, $D^*D^*$ and $D^*_sD^*$ molecular states. Our findings might be useful to get further insight in the flavor exotic sectors and can encourage the search for flavor-exotic mesons with e.g. double charm or double charm-strangeness in future experiments.

\section{Acknowledgments}  

T.B. acknowledges support from the DFG under Contract No. GRK683. This work is partly supported by DGICYT contract number
FIS2006-03438 and the Generalitat Valenciana in the Prometeo Program. We acknowledge the support of the European Community-Research Infrastructure
Integrating Activity
Study of Strongly Interacting Matter (acronym HadronPhysics2, Grant Agreement
n. 227431)
under the Seventh Framework Programme of EU.
\section{Appendix}
\subsection{Tree-level transition amplitudes of the four-vector contact
diagrams and of the $t$($u$)-channel vector-exchange diagrams 
for the different channels.}
\begin{table}[H]
\begin{center}
\small{\begin{tabular}{cc|cc|c}
%I=1/2,C=0,S=1
 $J$&Amplitude&Contact & V-exchange & $\sim$ Total\\
 \hline\hline
 $0$&$D_s^*\bar{D}^*\to D_s^*\bar{D}^*$&$2g^2$&$-\frac{g^2(p_1+p_3).(p_2+p_4)}{m_{J/\psi}^2}$&$0.23g^2$\T\B\\
 $0$&$D_s^*\bar{D}^*\to J/\psi K^*$&$-4g^2$&$\frac{g^2(p_1+p_4).(p_2+p_3)}{m_{D^*}^2}+\frac{g^2(p_1+p_3).(p_2+p_4)}{m_{D^*_s}^2}$&$3.6g^2$\T\B\\
$0$&$J/\psi K^*\to J/\psi K^*$&$0$&$0$&$0$\T\B\\
$1$&$D_s^*\bar{D}^*\to D_s^*\bar{D}^*$&$3g^2$&$-\frac{g^2(p_1+p_3).(p_2+p_4)}{m_{J/\psi}^2}$&$ 1.2g^2$\T\B\\
 $1$&$D_s^*\bar{D}^*\to J/\psi K^*$&$0$&$-\frac{g^2(p_1+p_4).(p_2+p_3)}{m_{D^*}^2}+\frac{g^2(p_1+p_3).(p_2+p_4)}{m_{D^*_s}^2}$&$-0.43g^2$\T\B\\
$1$&$J/\psi K^*\to J/\psi K^*$&$0$&$0$&$0$\T\B\\
$2$&$D_s^*\bar{D}^*\to D_s^*\bar{D}^*$&$-g^2$&$-\frac{g^2(p_1+p_3).(p_2+p_4)}{m_{J/\psi}^2}$&$-2.8g^2$\T\B\\
$2$&$D_s^*\bar{D}^*\to J/\psi K^*$&$2g^2$&$\frac{g^2(p_1+p_4).(p_2+p_3)}{m_{D^*}^2}+\frac{g^2(p_1+p_3).(p_2+p_4)}{m_{D^*_s}^2}$&$9.6g^2$\T\B\\
$2$&$J/\psi K^*\to J/\psi K^*$&$0$&$0$&$0$\T\B\\
\hline
\end{tabular}
\caption{Amplitudes for $C=0$, $S=1$ and $I=1/2$.\label{tab:p0}}}
\end{center}
\end{table}
\begin{table}[H]
\begin{center}
\small{\begin{tabular}{cc|cc|c}
%I=0,C=1,S=-1
 $J$&Amplitude&Contact & V-exchange& $\sim $ Total\\
 \hline\hline
$0$&$D^*\bar{K}^*\to D^*\bar{K}^*$&$4g^2$&$-\frac{g^2(p_1+p_4).(p_2+p_3)}{m_{D^*_s}^2}+\frac{1}{2}g^2(\frac{1}{m_\omega^2}-\frac{3}{m_\rho^2})(p_1+p_3).(p_2+p_4)$&$-9.9g^2$\T\B\\
$1$&$D^*\bar{K}^*\to D^*\bar{K}^*$&$0$&$\frac{g^2(p_1+p_4).(p_2+p_3)}{m_{D^*_s}^2}+\frac{1}{2}g^2(\frac{1}{m_\omega^2}-\frac{3}{m_\rho^2})(p_1+p_3).(p_2+p_4)$&$-10.2g^2$\T\B\\
$2$&$D^*\bar{K}^*\to D^*\bar{K}^*$&$-2g^2$&$-\frac{g^2(p_1+p_4).(p_2+p_3)}{m_{D^*_s}^2}+\frac{1}{2}g^2(\frac{1}{m_\omega^2}-\frac{3}{m_\rho^2})(p_1+p_3).(p_2+p_4)$&$-15.9g^2$\T\B\\
\hline
\end{tabular}
\caption{Amplitudes for $C=1$, $S=-1$ and $I=0$.\label{tab:p1}}}
\end{center}
\end{table}
\begin{table}[H]
\begin{center}
\small{\begin{tabular}{cc|cc|c}
%I=1,C=1,S=-1
 $J$&Amplitude&Contact & V-exchange & $\sim$ Total\\
 \hline\hline
$0$&$D^*\bar{K}^*\to D^*\bar{K}^*$&$-4g^2$&$\frac{g^2(p_1+p_4).(p_2+p_3)}{m_{D^*_s}^2}+\frac{g^2}{2}(\frac{1}{m_\omega^2}+\frac{1}{m_\rho^2})(p_1+p_3).(p_2+p_4)$&$9.7g^2$\T\B\\
$1$&$D^*\bar{K}^*\to D^*\bar{K}^*$&$0$&$-\frac{g^2(p_1+p_4).(p_2+p_3)}{m_{D^*_s}^2}+\frac{g^2}{2}(\frac{1}{m_\omega^2}+\frac{1}{m_\rho^2})(p_1+p_3).(p_2+p_4)$&$9.9g^2$\T\B\\
$2$&$D^*\bar{K}^*\to D^*\bar{K}^*$&$2g^2$&$\frac{g^2(p_1+p_4).(p_2+p_3)}{m_{D^*_s}^2}+\frac{g^2}{2}(\frac{1}{m_\omega^2}+\frac{1}{m_\rho^2})(p_1+p_3).(p_2+p_4)$&$15.7g^2$\T\B\\
\hline
\end{tabular}
\caption{Amplitudes for $C=1$, $S=-1$ and $I=1$.\label{tab:p2}}}
\end{center}
\end{table}
\begin{table}[H]
\begin{center}
\small{\begin{tabular}{cc|cc|c}
%I=0,C=1,S=1
 $J$&Amplitude&Contact & V-exchange& $\sim$ Total\\
 \hline\hline
$0$&$D^*K^*\to D^*K^*$&$4g^2$&$-\frac{g^2}{2}(\frac{3}{m_\rho^2}+\frac{1}{m_\omega^2}) (p_1+p_3).(p_2+p_4)$&$-19.8g^2$\T\B\\
$0$&$D^*K^*\to D_s^*\omega$&$-4g^2 $&$\frac{g^2(p_1+p_4).(p_2+p_3)}{m_{D^*_s}^2}+\frac{g^2(p_1+p_3).(p_2+p_4)}{m_{K^*}^2}$&$6.8g^2$\T\B\\
$0$&$D^*K^*\to D_s^*\phi$&$2\sqrt{2}g^2$&$-\frac{\sqrt{2}g^2(p_1+p_3).(p_2+p_4)}{m_{K^*}^2}$&$-9.2g^2$\T\B\\
$0$&$D_s^*\omega\to D_s^*\omega$&$0$&$0$&$0$\T\B\\
$0$&$D^*_s\omega\to D_s^*\phi$&$0$&$0$&$0$\T\B\\
$0$&$D_s^*\phi\to D_s^*\phi$&$-2g^2$&$\frac{g^2(p_1+p_4).(p_2+p_3)}{m_{D^*_s}^2}$&$0.20g^2$\T\B\\
$1$&$D^*K^*\to D^*K^*$&$6g^2$&$-\frac{g^2}{2}(\frac{3}{m_\rho^2}+\frac{1}{m_\omega^2}) (p_1+p_3).(p_2+p_4)$&$-17.7g^2$\T\B\\
$1$&$D^*K^*\to D_s^*\omega$&$0 $&$-\frac{g^2(p_1+p_4).(p_2+p_3)}{m_{D^*_s}^2}+\frac{g^2(p_1+p_3).(p_2+p_4)}{m_{K^*}^2}$&$6.6g^2$\T\B\\
$1$&$D^*K^*\to D_s^*\phi$&$3\sqrt{2}g^2 $&$-\frac{\sqrt{2}g^2(p_1+p_3).(p_2+p_4)}{m_{K^*}^2}$&$-7.8g^2$\T\B\\
$1$&$D_s^*\omega\to D_s^*\omega$&$0$&$0$&$0$\T\B\\
$1$&$D_s^*\omega\to D_s^*\phi$&$0$&$0$&$0$\T\B\\
$1$&$D_s^*\phi\to D_s^*\phi$&$3g^2$&$-\frac{g^2(p_1+p_4).(p_2+p_3)}{m_{D^*_s}^2}$&$0.8g^2$\T\B\\
$2$&$D^*K^*\to D^*K^*$&$-2g^2$&$-\frac{g^2}{2}(\frac{3}{m_\rho^2}+\frac{1}{m_\omega^2}) (p_1+p_3).(p_2+p_4)$&$-25.8g^2$\T\B\\
$2$&$D^*K^*\to D_s^*\omega$&$2g^2 $&$\frac{g^2(p_1+p_4).(p_2+p_3)}{m_{D^*_s}^2}+\frac{g^2(p_1+p_3).(p_2+p_4)}{m_{K^*}^2}$&$12.8g^2$\T\B\\
$2$&$D^*K^*\to D_s^*\phi$&$-\sqrt{2}g^2 $&$-\frac{\sqrt{2}g^2(p_1+p_3).(p_2+p_4)}{m_{K^*}^2}$&$-13.5g^2$\T\B\\
$2$&$D_s^*\omega\to D_s^*\omega$&$0$&$0$&$0$\T\B\\
$2$&$D_s^*\omega\to D^*_s\phi$&$0$&$0$&$0$\T\B\\
$2$&$D_s^*\phi\to D_s^*\phi$&$g^2$&$\frac{g^2(p_1+p_4).(p_2+p_3)}{m_{D^*_s}^2}$&$3.2g^2$\T\B\\
\hline
\end{tabular}
\caption{Amplitudes for $C=1$, $S=1$ and $I=0$.\label{tab:p3}}}
\end{center}
\end{table}
\begin{table}[H]
\begin{center}
\small{\begin{tabular}{cc|cc|c}
%I=1,C=1,S=1
 $J$&Amplitude&Contact & V-exchange& $\sim$ Total\\
 \hline\hline
$0$&$D^*K^*\to D^*K^*$&$0$&$\frac{g^2}{2}(\frac{1}{m_\rho^2}-\frac{1}{m_\omega^2})(p_1+p_3).(p_2+p_4)$&$0.11g^2$\T\B\\
$0$&$D^*K^*\to D^*_s\rho$&$4g^2$&$-\frac{g^2(p_1+p_4)(p_2+p_3)}{m_{D^*}^2}-\frac{g^2(p_1+p_3).(p_2+p_4)}{m_{K^*}^2}$&$-6.8g^2$\T\B\\
$0$&$D_s^*\rho\to D^*_s\rho$&$0$&$0$&$0$\T\B\\
$1$&$D^*K^*\to D^*K^*$&$0$&$\frac{g^2}{2}(\frac{1}{m_\rho^2}-\frac{1}{m_\omega^2})(p_1+p_3).(p_2+p_4)$&$0.11g^2$\T\B\\
$1$&$D^*K^*\to D^*_s\rho$&$0$&$\frac{g^2(p_1+p_4)(p_2+p_3)}{m_{D^*}^2}-\frac{g^2(p_1+p_3).(p_2+p_4)}{m_{K^*}^2}$&$-6.6g^2$\T\B\\
$1$&$D_s^*\rho\to D^*_s\rho$&$0$&$0$&$0$\T\B\\
$2$&$D^*K^*\to D^*K^*$&$0$&$\frac{g^2}{2}(\frac{1}{m_\rho^2}-\frac{1}{m_\omega^2})(p_1+p_3).(p_2+p_4)$&$0.11g^2$\T\B\\
$2$&$D^*K^*\to D^*_s\rho$&$-2g^2$&$-\frac{g^2(p_1+p_4)(p_2+p_3)}{m_{D^*}^2}-\frac{g^2(p_1+p_3).(p_2+p_4)}{m_{K^*}^2}$&$-12.8g^2$\T\B\\
$2$&$D_s^*\rho\to D^*_s\rho$&$0$&$0$&$0$\T\B\\
\hline
\end{tabular}
\caption{Amplitudes for $C=1$, $S=1$ and $I=1$.\label{tab:p4}}}
\end{center}
\end{table}
\begin{table}[H]
\begin{center}
\small{\begin{tabular}{cc|cc|c}
%I=1/2,C=1,S=2
 $J$&Amplitude&Contact & V-exchange& $\sim$ Total\\
 \hline\hline
$0$&$D^*_sK^*\to D^*_sK^*$&$-4g^2$&$\frac{g^2(p_1+p_4)(p_2+p_3)}{m_{D^*}^2}+\frac{g^2(p_1+p_3).(p_2+p_4)}{m_\phi^2}$&$5.5g^2$\T\B\\
$1$&$D^*_sK^*\to D^*_sK^*$&$0$&$-\frac{g^2(p_1+p_4)(p_2+p_3)}{m_{D^*}^2}+\frac{g^2(p_1+p_3).(p_2+p_4)}{m_\phi^2}$&$5.0g^2$\T\B\\
$2$&$D^*_sK^*\to D^*_sK^*$&$2g^2$&$\frac{g^2(p_1+p_4)(p_2+p_3)}{m_{D^*}^2}+\frac{g^2(p_1+p_3).(p_2+p_4)}{m_\phi^2}$&$11.5g^2$\T\B\\
\hline
\end{tabular}
\caption{Amplitudes for $C=1$, $S=2$ and $I=1/2$.\label{tab:p5}}}
\end{center}
\end{table}
\begin{table}[H]
\begin{center}
\small{\begin{tabular}{cc|cc|c}
%I=0,C=2,S=0
 $J$&Amplitude&Contact & V-exchange& $\sim$ Total\\
 \hline\hline
$0$&$D^*D^*\to D^*D^*$&$0$&$0$&$0$\T\B\\
$1$&$D^*D^*\to D^*D^*$&$0$&$\frac{1}{4}g^2(\frac{2}{m_{J/\psi}^2}+\frac{1}{m_\omega^2}-\frac{3}{m_\rho^2})\lbrace(p_1+p_4).(p_2+p_3)$ &$-25.4g^2$ \T\B\\
& & &\hspace{3.7cm}$+(p_1+p_3).(p_2+p_4)\rbrace$&\T\B\\
$2$&$D^*D^*\to D^*D^*$&$0$&$0$&$0$\T\B\\
\hline
\end{tabular}
\caption{Amplitudes for $C=2$, $S=0$ and $I=0$.\label{tab:p6}}}
\end{center}
\end{table}
\begin{table}[H]
\begin{center}
\small{\begin{tabular}{cc|cc|c}
%I=0,C=2,S=0
 $J$&Amplitude&Contact & V-exchange& $\sim$ Total\\
 \hline\hline
$0$&$D^*D^*\to D^*D^*$&$-4g^2$&$\frac{1}{4}g^2(\frac{2}{m_{J/\psi}^2}+\frac{1}{m_\omega^2}+\frac{1}{m_\rho^2})\lbrace(p_1+p_4).(p_2+p_3)$ &$24.3g^2$ \T\B\\
 & & &\hspace{3.7cm}$+(p_1+p_3).(p_2+p_4)\rbrace$&\T\B\\
$1$&$D^*D^*\to D^*D^*$&$0$&$0$&$0$\T\B\\
$2$&$D^*D^*\to D^*D^*$&$2g^2$&$\frac{1}{4}g^2(\frac{2}{m_{J/\psi}^2}+\frac{1}{m_\omega^2}+\frac{1}{m_\rho^2})\lbrace(p_1+p_4).(p_2+p_3)$& $30.3g^2$\T\B\\
 & & &\hspace{3.7cm}$+(p_1+p_3).(p_2+p_4)\rbrace$&\T\B\\
\hline
\end{tabular}
\caption{Amplitudes for $C=2$, $S=0$ and $I=1$.\label{tab:p7}}}
\end{center}
\end{table}
\begin{table}[htpb]
\begin{center}
\small{\begin{tabular}{cc|cc|c}
%I=1/2,C=2,S=1
 $J$&Amplitude&Contact & V-exchange& $\sim$ Total\\
 \hline\hline
$0$&$D^*_sD^*\to D^*_sD^*$&$-4g^2$&$\frac{g^2(p_1+p_4).(p_2+p_3)}{m_{K^*}^2}+\frac{g^2(p_1+p_3).(p_2+p_4)}{m_{J/\psi^2}}$&$19.0g^2$\T\B\\
$1$&$D^*_sD^*\to D^*_sD^*$&$0$&$-\frac{g^2(p_1+p_4).(p_2+p_3)}{m_{K^*}^2}+\frac{g^2(p_1+p_3).(p_2+p_4)}{m_{J/\psi^2}}$&$-19.5g^2$\T\B\\
$2$&$D^*_sD^*\to D^*_sD^*$&$2g^2$&$\frac{g^2(p_1+p_4).(p_2+p_3)}{m_{K^*}^2}+\frac{g^2(p_1+p_3).(p_2+p_4)}{m_{J/\psi^2}}$&$25.0g^2$\T\B\\
\hline
\end{tabular}
\caption{Amplitudes for $C=2$, $S=1$ and $I=1/2$.\label{tab:p8}}}
\end{center}
\end{table}
\begin{table}[htpb]
\begin{center}
\small{\begin{tabular}{cc|cc|c}
%I=0,C=2,S=2
 $J$&Amplitude&Contact & V-exchange & $\sim$ Total\\
 \hline \hline
$0$&$D^*_sD^*_s\to D^*_sD^*_s$&$-4g^2$&$\frac{g^2}{2}(\frac{1}{m_{J/\psi}^2}+\frac{1}{m_\phi^2})\lbrace(p_1+p_4).(p_2+p_3)+(p_1+p_3).(p_2+p_4)\rbrace$&$15.0g^2$\T\B\\
$1$&$D^*_sD^*_s\to D^*_sD^*_s$&$0$&$0$&$0$\T\B\\
$2$&$D^*_sD^*_s\to D^*_sD^*_s$&$2g^2$&$\frac{g^2}{2}(\frac{1}{m_{J/\psi}^2}+\frac{1}{m_\phi^2})\lbrace(p_1+p_4).(p_2+p_3)+(p_1+p_3).(p_2+p_4)\rbrace$&$21.0g^2$\T\B\\
\hline
\end{tabular}
\caption{Amplitudes for $C=2$, $S=2$ and $I=0$.\label{tab:p9}}}
\end{center}
\end{table}

\subsection{Box diagrams}
\subsubsection{$D^* K^*\to D^*K^*$ box diagram with $m_1=\pi$, $m_2=D$, $m_3=\pi$ and $m_4=K$ }
\begin{eqnarray}\label{eq:boxdk}
V_{D^*K^*}(s)& = &\int^{q_{max}}_0 dq 
 \frac{q^6 }{ \omega^3 \omega_K \omega_D}\frac{1}{(-k_3^0 + \omega+\omega_D  - 
        i \eps)^2 }\frac{1}{(-\sqrt{s} + \omega_D+\omega_K- 
       i \eps) }\nonumber\\&\times&\frac{1}{(-k_4^0 + \omega+\omega_K - 
        i\eps)^2 }\frac{1}{(k_3^0 + \omega+\omega_D)^2 }\frac{1}{(\sqrt{s} + \omega_D+\omega_K)}\nonumber\\&\times&\frac{1}{(k_4^0 + \omega+\omega_K)^2 }\times\frac{ g^4}{15 \pi^2} P(s)\ ,
 \end{eqnarray}
 with
 \begin{eqnarray}
P(s)&=& -2 (2 k_3^0 \sqrt{s} (\omega + \omega_D) \omega_K (4 \omega^4 - s (4 \omega^2 + 3 \omega \omega_D + \omega_D^2)+ 
        4 \omega^2 (\omega_D + \omega_K)^2 \nonumber\\& +& 3 \omega \omega_D (\omega_D + \omega_K)^2 + 
        \omega_D^2 (\omega_D + \omega_K)^2 + 4 \omega^3 (\omega_D + 2 \omega_K)) \nonumber\\&+& 
     2 (k_3^{0})^3 \sqrt{s}
       \omega_K(-4 \omega^3 - \omega_D (-s + (\omega_D + \omega_K)^2)) \nonumber\\&+& 
     (k_3^0)^4 (2 \omega^3 (\omega_D + \omega_K) + 
        \omega_D \omega_K (-s + (\omega_D + \omega_K)^2))\nonumber\\& - &
     (k_3^0)^2 (4 \omega^5 (\omega_D + \omega_K) + 
        8 \omega^4 (\omega_D + \omega_K)^2 \nonumber\\&+& 
        4 \omega^3 (\omega_D^3 - 3 s \omega_K + 6 \omega_D^2 \omega_K + 
           6 \omega_D \omega_K^2 + \omega_K^3) \nonumber\\&+& 
        14 \omega^2 \omega_D\omega_K(-s + (\omega_D + \omega_K)^2) + 
        4 \omega\omega_D\omega_K (\omega_D+\omega_K) \nonumber\\&\times&(-s + (\omega_D+\omega_K)^2) + 
        \omega_D \omega_K(s^2 + (\omega_D+\omega_K)^2 (\omega_D^2 + \omega_K^2) \nonumber\\&-& 
           2 s (\omega_D^2 + \omega_D\omega_K + \omega_K^2))) + (\omega + 
        \omega_D)^2 (2 \omega^5 (\omega_D+\omega_K) \nonumber\\&+& 
        4 \omega^4 (\omega_D^2 + 3 \omega_D\omega_K + 2 \omega_K^2) + 
        2 \omega^3 (\omega_D^3 - 2 s \omega_K + 7 \omega_D^2 \omega_K + 
           12 \omega_D \omega_K^2 + 6 \omega_K^3) \nonumber\\&+& 
        \omega_D\omega_K (s - \omega_K^2) (s - (\omega_D+\omega_K)^2) + 
        \omega^2 \omega_K (5 \omega_D + 8 \omega_K) (-s + (\omega_D + \omega_K)^2) \nonumber\\&+& 
        2 \omega\omega_K (s^2 + \omega_K (\omega_D+\omega_K)^2 (2 \omega_D+\omega_K) - 
           s (\omega_D^2 + 4 \omega_D\omega_K + 2 \omega_K^2))))\ .
	   \end{eqnarray}
	   Where $\omega=\sqrt{q^2+m_\pi^2}$, $\omega_D=\sqrt{q^2+m_D^2}$, $\omega_K=\sqrt{q^2+m_K^2}$,  $k^0_3=\frac{s+m^2_{D^*}-m^2_{K^*}}{2\sqrt{s}}$ and $k^0_4=\frac{s+m^2_{K^*}-m^2_{D^*}}{2\sqrt{s}}$. After projecting in spin and isospin, the potential is
	
	   \begin{eqnarray}
	   V_{D^*K^* }^{I=0,J=0}(s)&=&\frac{9}{4}\,5\,V_{D^*K^*}(s)\nonumber\\
	   V_{D^*K^* }^{I=0,J=2}(s)&=&\frac{9}{4}\,2\,V_{D^*K^* }(s)\nonumber\\
	   V_{D^*K^* }^{I=1,J=0}(s)&=&\frac{1}{4}\,5\,V_{D^*K^*}(s)\nonumber\\
	   V_{D^*K^* }^{I=1,J=2}(s)&=&\frac{1}{4}\,2\,V_{D^*K^* }(s)\ .
	   \end{eqnarray}

	   \subsubsection{$D^* K^*\to D_s^*\phi$ box diagram with $m_1=\pi$, $m_2=D$, $m_3=K$ and $m_4=K$ }
\begin{eqnarray}\label{eq:boxdkdsphi}
V_{D^*K^*\to D^*_s\phi}(s)& = &\int^{q_{max}}_0 dq 
 \frac{q^6}{\omega \omega_D\omega_K^2 }\frac{1}{(-k_1^0 + \omega+\omega_D - i\eps) }\frac{1}{ (k_1^0- k_3^0 +\omega + \omega_K -i\eps)}\nonumber\\&\times&\frac{1}{ (-k_1^0 + k_3^0 + 
     \omega+\omega_K- i\eps) }\frac{1}{(-k_2^0 + \omega+\omega_K - i\eps) } \nonumber\\&\times&\frac{1}{ (-k_3^0 + \omega_D+\omega_K - i\eps) }\frac{1}{ (-k_4^0 + 2 \omega_K - i\eps)}\nonumber\\&\times&\frac{1}{ (-\sqrt{s} + \omega_D+\omega_K - i\eps) }\frac{1}{(k_1^0 +\omega+\omega_D)}\nonumber\\&\times&\frac{1}{(k_3^0+\omega_D + 
     \omega_K)}\frac{1}{(k_2^0 + 
    \omega+\omega_K)}\nonumber\\&\times&\frac{1}{(\sqrt{s} +\omega_D+ 
     \omega_k)}\frac{1}{ (k^0_4 + 2 \omega_k)}\times \frac{4 g^4}{15\pi^2} P(s)\ ,
 \end{eqnarray}
 with
 \begin{eqnarray}
P(s)&=& -\omega_K(2 (k_1^0)^3 \omega (-(k_3^0)^2 \sqrt{s} \omega_K + (k_3^0)^3 (\omega_D+\omega_K) \nonumber\\&+& 
      \sqrt{s} (\omega_D+\omega_K) (s - \omega_D^2 - 3 \omega_D\omega_K - 
         4 \omega_K^2) \nonumber\\&-&
      k_3^0 (\omega_D^3 + s \omega_K + 4 \omega_D^2 \omega_K + 7 \omega_D\omega_K^2 + 
         4 \omega_K^3)) \nonumber\\&- &
   (k_1^0)^4 \omega (-2 k_3^0 \sqrt{s}\omega_K + 
      (k_3^0)^2 (\omega_D+\omega_K) \nonumber\\&+& (\omega_D+\omega_K) (s - 
         2 (\omega_D^2 + 3 \omega_D\omega_K + 2 \omega_K^2))) \nonumber\\&+ &
   2 k_1^0 (k_3^0 + \sqrt{s}) \omega ((k_3^0)^3 \sqrt{s}\omega_K \nonumber\\&-& 
      (k_3^0)^2 (\omega_D^3 + 2 s \omega_K + 2 \omega_D^2 \omega_K + 
         2 \omega_D\omega_K^2 + \omega_K^3 + \omega^2 (\omega_D+\omega_K) \nonumber\\&+& 
         2 \omega (\omega_D+\omega_K)^2) + 
      k_3^0\sqrt{s}(\omega_D^3 + s \omega_K + 4 \omega_D^2 \omega_K + 4 \omega_D\omega_K^2 - 
         2\omega_K^3 \nonumber\\&+& \omega^2 (\omega_D + 2 \omega_K) + 
         2 \omega (\omega_D^2 + 4 \omega_D\omega_K + 2 \omega_K^2)) \nonumber\\&+& (\omega_D+\omega_K) (\omega_D^4 + 3 \omega_D^3 \omega_K + 4 \omega_D^2 \omega_K^2 \nonumber\\&+& 
         4 \omega_D\omega_K^3 + 4 \omega_K^4 + 
         \omega^2 (\omega_D^2 + 3 \omega_D\omega_K + 4 \omega_K^2) \nonumber\\&+& 
         2 \omega (\omega_D^3 + 3 \omega_D^2 \omega_K + 4 \omega_D\omega_K^2 + 
            4 \omega_K^3) \nonumber\\&-& 
         s (\omega^2 + \omega_D^2 + \omega_D\omega_K + \omega_K^2 + 
            2 \omega (\omega_D+\omega_K)))) \nonumber\\&-& 
   (k_1^0)^2 \omega (2 (k_3^0)^3 \sqrt{s}\omega_K + (k_3^0)^4 (\omega_D+\omega_K)\nonumber\\& -& 
      2 (k_3^0)^2 (\omega_D^3 + \omega_D^2\omega_K+ 3 \omega_D\omega_K^2 + 
         \omega^2 (\omega_D+\omega_K) + 3 \omega_K (s + \omega_K^2) \nonumber\\&+& 
         \omega (\omega_D^2 + 3 \omega_D\omega_K+ 2 \omega_K^2)) \nonumber\\&+& 
      2 k_3^0 \sqrt{s}
        \omega_K (s + 
         2 (\omega^2 - \omega_D^2 - 4 \omega_D\omega_K - 3 \omega_K^2 + 
            \omega (\omega_D + 2 \omega_K))) \nonumber\\&+& (\omega_D+\omega_K) (s^2 - 
         2 s ( \omega^2 +\omega_D^2 + 3 \omega_K^2 + 
            \omega (\omega_D + 2 \omega_K)) \nonumber\\&+& 
         2 (\omega_D+\omega_K) (\omega_D^3 + 2 \omega_D^2 \omega_K + 
            5 \omega_D\omega_K^2 + 4 \omega_K^3 \nonumber\\&+& 
            2 \omega^2 (\omega_D + 2 \omega_K) + 
            2 \omega (\omega_D + 2 \omega_K)^2))) \nonumber\\&+& (\omega + 
      \omega_D) (2 (k_3^0)^3 \sqrt{s}
        \omega_K (s - \omega^2 - (\omega_D+\omega_K)^2 \nonumber\\&-& 
         \omega (\omega_D+ 2 \omega_K)) + 
      (k_3^0)^4 (\omega^2 (\omega_D+\omega_K) \nonumber\\&+& 
         \omega (\omega_D^2 + 3 \omega_D\omega_K + 2 \omega_K^2) + 
         \omega_K (-s + (\omega_D+\omega_K)^2)) \nonumber\\&+& 
      2 k_3^0 \sqrt{s}
       \omega_K (\omega^4 + (\omega_D+\omega_K)^4 + \omega^3 (\omega_D + 4 \omega_K) \nonumber\\&+&
          \omega^2 (\omega_D^2 + 4 \omega_D\omega_K + 6 \omega_K^2) + 
         \omega(\omega_D^3 + 4 \omega_D^2 \omega_K + 6 \omega_D\omega_K^2 + 
            4 \omega_K^3)\nonumber\\& -& 
         s (\omega^2 + (\omega_D+ \omega_K)^2 + \omega (\omega_D+ 2 \omega_K))) - 
      (k_3^0)^2 (\omega^4 (\omega_D+\omega_K) \nonumber\\&+& 
         \omega^3 (\omega_D^2 + 5 \omega_D\omega_K + 4 \omega_K^2) + 
         \omega^2 (\omega_D^3 - 2 s \omega_K + 9 \omega_D^2 \omega_K + 
            18 \omega_D \omega_K^2 + 10 \omega_K^3) \nonumber\\&+& 
         \omega (\omega_D^4 + 2 s \omega_D\omega_K + 5 \omega_D^3 \omega_K - 
            4 s \omega_K^2 + 18 \omega_D^2 \omega_K^2 + 26 \omega_D\omega_K^3 + 
            12 \omega_K^4) \nonumber\\&+& 
         \omega_K (s^2 - 
            2 s (\omega_D^2 + 2 \omega_D\omega_K + 3 \omega_K^2) + (\omega_D + 
               \omega_K)^2 (\omega_D^2 + 2 \omega_D\omega_K + 5 \omega_K^2))) \nonumber\\&+& (\omega + \omega_K) (\omega_D+\omega_K) (s^2 (\omega+\omega_D+\omega_K) \nonumber\\&+& 
         2 (\omega+\omega_K) (\omega_D+\omega_K) (2 \omega_K (\omega_D+\omega_K)^2 \nonumber\\&+&
             \omega^2 (\omega_D + 2 \omega_K) + \omega (\omega_D + 2 \omega_K)^2) \nonumber\\&-& 
         s (\omega^3 + \omega_D^3 + 3 \omega_D^2 \omega_K + 7 \omega_D\omega_K^2 + 
            5 \omega_K^3\nonumber\\& +& \omega^2 (\omega_D + 3 \omega_K) + 
            \omega (\omega_D^2 + 7\omega_D\omega_K + 7 \omega_K^2)))))\ .
	   \end{eqnarray}
	   Where $\omega=\sqrt{q^2+m_\pi^2}$, $\omega_D=\sqrt{q^2+m_D^2}$, $\omega_K=\sqrt{q^2+m_K^2}$,  $k^0_1=\frac{s+m^2_{D^*}-m^2_{K^*}}{2\sqrt{s}}$, $k^0_2=\frac{s+m^2_{K^*}-m^2_{D^*}}{2\sqrt{s}}$ and $k^0_3=\frac{s+m^2_{D^*_s}-m^2_{\phi}}{2\sqrt{s}}$. And its spin-isospin projection is
	   
	   \begin{eqnarray}
	   V_{D^*K^* \to D_s^*\phi}^{I=0,J=0}(s)&=&\frac{3}{\sqrt{2}}\,5\,V_{D^*K^*\to D_s^*\phi}(s)\nonumber\\
	   V_{D^*K^*\to D_s^*\phi }^{I=0,J=2}(s)&=&\frac{3}{\sqrt{2}}\,2\,V_{D^*K^*\to D_s^*\phi }(s)\ .
	   \end{eqnarray}

	   \subsubsection{$D^*_s \phi\to D^*_s\phi$ box diagram with $m_1=K$, $m_2=D$, $m_3=K$ and $m_4=K$ }
\begin{eqnarray}\label{eq:boxdsphi}
V_{D^*_s\phi }(s)&=&\int^{q_{max}}_0 dq 
 \frac{q^6 }{ \omega_D\omega_K^3 }\frac{1}{(-k^0_3 +\omega_D+\omega_K- i\eps)^2 }\frac{1}{(-\sqrt{s} +\omega_D+ \omega_K-  i\eps) }\nonumber\\&\times&\frac{1}{(-k^0_4 + 2 \omega_K - 
    i\eps)^2}\frac{1}{(\sqrt{s} + \omega_D+\omega_K) }\frac{1}{ (k^0_4 + 2 \omega_K)^2 }\nonumber\\&\times&\frac{1}{(k^0_3 + \omega_D+\omega_K)^2 }\times\frac{ g^4}{15 \pi^2} P(s)\ ,
 \end{eqnarray}
 with
 \begin{eqnarray}
P(s)&=&2 ((k_3^0)^4 (s \omega_D - \omega_D^3 - 2\omega_D^2 \omega_K - 3 \omega_D\omega_K^2 - 
      2 \omega_K^3) \nonumber\\&+& 
   2 (k_3^0)^3 \sqrt{s} (-s \omega_D + \omega_D^3 + 2 \omega_D^2 \omega_K + \omega_D\omega_K^2 + 
      4\omega_K^3) \nonumber\\&-& 
   2 k^0_3 \sqrt{s}(\omega_D+\omega_K) (\omega_D^4 + 5 \omega_D^3 \omega_K +
      11 \omega_D^2 \omega_K^2 \nonumber\\&+& 15 \omega_D\omega_K^3 + 16 \omega_K^4 - 
      s (\omega_D^2 + 3 \omega_D\omega_K + 4 \omega_K^2)) \nonumber\\&+&  
   (k_3^0)^2 (s^2 \omega_D+\omega_D^5 + 6\omega_D^4 \omega_K +
      32 \omega_D^3 \omega_K^2 + 74 \omega_D^2 \omega_K^3 \nonumber\\&+& 63 \omega_D\omega_K^4 + 
      16 \omega_K^5 -
      2 s (\omega_D^3 + 3 \omega_D^2 \omega_K + 10 \omega_D\omega_K^2 + 
         6 \omega_K^3)) \nonumber\\&-& (\omega_D+\omega_K)^2 (s^2 (\omega_D + 2 \omega_K) + 
      4 \omega_K^2 (3 \omega_D^3 + 12 \omega_D^2 \omega_K \nonumber\\&+& 17 \omega_D\omega_K^2 + 
         8 \omega_K^3) - 
      s (\omega_D^3 + 4 \omega_D^2 \omega_K + 15 \omega_D\omega_K^2 + 16 \omega_K^3)))\ .
      \end{eqnarray}
	    Where $\omega=\sqrt{q^2+m_\pi^2}$, $\omega_D=\sqrt{q^2+m_D^2}$, $\omega_K=\sqrt{q^2+m_K^2}$, $k^0_3=\frac{s+m^2_{D_s^*}-m^2_{\phi}}{2\sqrt{s}}$, $k^0_4=\frac{s+m^2_{\phi}-m^2_{D^*_s}}{2\sqrt{s}}$ and $\eps=1$ MeV. Finally, we project it in spin and isospin
	      \begin{eqnarray}
	   V_{D^*_s\phi }^{I=0,J=0}(s)&=&2\times5\,V_{D^*_s\phi }(s)\nonumber\\
	   V_{D^*_s\phi }^{I=0,J=2}(s)&=&2\times2\,V_{D^*_s\phi }(s)\ .
	   \end{eqnarray}
\subsubsection{$D^* \bar{K}^*\to D^*\bar{K}^*$ box diagram with $m_1=\pi$, $m_2=D$, $m_3=\pi$ and $m_4=\bar{K}$}	   
The potential is the same than that given by Eq. (\ref{eq:boxdk}) with:

	   \begin{eqnarray}
	   V_{D^*\bar{K}^* }^{I=0,J=0}(s)&=&\frac{9}{4}\,5\,V_{D^*K^*}(s)\nonumber\\
	   V_{D^*\bar{K}^* }^{I=0,J=2}(s)&=&\frac{9}{4}\,2\,V_{D^*K^* }(s)\nonumber\\
	   V_{D^*\bar{K}^* }^{I=1,J=0}(s)&=&\frac{1}{4}\,5\,V_{D^*K^*}(s)\nonumber\\
	   V_{D^*\bar{K}^* }^{I=1,J=2}(s)&=&\frac{1}{4}\,2\,V_{D^*K^* }(s)\ .
	   \end{eqnarray}


\begin{thebibliography}{99}

%\cite{Kubota:1994gn}
\bibitem{Kubota}
  Y.~Kubota {\it et al.}  [CLEO Collaboration],
  %``Observation of a new charmed strange meson,''
  Phys.\ Rev.\ Lett.\  {\bf 72}, 1972 (1994)
 % [arXiv:hep-ph/9403325].
  %%CITATION = PRLTA,72,1972;%%
  

\bibitem{Godfrey2}
  S.~Godfrey and R.~Kokoski,
  %``The Properties of p Wave Mesons with One Heavy Quark,''
  Phys.\ Rev.\  D {\bf 43}, 1679 (1991).
  %%CITATION = PHRVA,D43,1679;%%
  %\cite{Godfrey:1985xj}
  
%\cite{Besson:2003cp}
\bibitem{Besson}
  D.~Besson {\it et al.}  [CLEO Collaboration],
  %``Observation of a narrow resonance of mass 2.46-GeV/c**2 decaying to  D/s*+
  %pi0 and confirmation of the D/sJ*(2317) state,''
  Phys.\ Rev.\  D {\bf 68}, 032002 (2003)
  [Erratum-ibid.\  D {\bf 75}, 119908 (2007)]
%  [arXiv:hep-ex/0305100].
  %%CITATION = PHRVA,D68,032002;%%
%\cite{Aubert:2003fg}
\bibitem{Aubert}
  B.~Aubert {\it et al.}  [BABAR Collaboration],
  %``Observation of a narrow meson decaying to $D_s^+ \pi^0$ at a mass of
  %2.32-GeV/c$^2$,''
  Phys.\ Rev.\ Lett.\  {\bf 90}, 242001 (2003)
%  [arXiv:hep-ex/0304021].
  %%CITATION = PRLTA,90,242001;%%
  
  \bibitem{Godfrey1}
  S.~Godfrey and N.~Isgur,
  %``Mesons In A Relativized Quark Model With Chromodynamics,''
  Phys.\ Rev.\  D {\bf 32}, 189 (1985).
  %%CITATION = PHRVA,D32,189;%%
%\cite{Godfrey:1986wj}
%\cite{Isgur:1991wq}

\bibitem{Pierro}
  M.~Di Pierro and E.~Eichten,
  %``Excited heavy-light systems and hadronic transitions,''
  Phys.\ Rev.\  D {\bf 64}, 114004 (2001)
  [arXiv:hep-ph/0104208].
  %%CITATION = PHRVA,D64,114004;%%

 %\cite{vanBeveren:2003kd}
\bibitem{vanBeveren}
  E.~van Beveren and G.~Rupp,
  %``Observed D/s(2317) and tentative D(2030) as the charmed cousins of the
  %light scalar nonet,''
  Phys.\ Rev.\ Lett.\  {\bf 91}, 012003 (2003)
  [arXiv:hep-ph/0305035].
  %%CITATION = PRLTA,91,012003;%%
 %\cite{Hwang:2004cd}
\bibitem{Hwang}
  D.~S.~Hwang and D.~W.~Kim,
  %``Mass of D*_sJ(2317) and Coupled Channel Effect,''
  Phys.\ Lett.\  B {\bf 601}, 137 (2004)
%  [arXiv:hep-ph/0408154].
  %%CITATION = PHLTA,B601,137;%%
  
  %\cite{Simonov:2004ar}
\bibitem{Simonov}
  Yu.~A.~Simonov and J.~A.~Tjon,
  %``The coupled-channel analysis of the D and D_s mesons,''
  Phys.\ Rev.\  D {\bf 70}, 114013 (2004)
%  [arXiv:hep-ph/0409361].
  %%CITATION = PHRVA,D70,114013;%%
  %\cite{Becirevic:2004uv}
\bibitem{Becirevic}
  D.~Becirevic, S.~Fajfer and S.~Prelovsek,
  %``On the mass differences between the scalar and pseudoscalar heavy-light
  %mesons,''
  Phys.\ Lett.\  B {\bf 599}, 55 (2004)
%  [arXiv:hep-ph/0406296].
  %%CITATION = PHLTA,B599,55;%%
  \bibitem{Eichten}
  \textit{in Proceedings of the 3rd International Workshop on Heavy Quarkonium, October 12-15, 2004 (IHEP, Beijing, 2004)}. ArXiv:hep-ph/0412158.
  \bibitem{Rupp}
  G. Rupp, F. Kleefeld, and E. van Beveren, AIP Conf. Proc. {\bf 756}, 360 (2005).

%\cite{Gamermann:2006nm}
\bibitem{Gamermann1}
  D.~Gamermann, E.~Oset, D.~Strottman and M.~J.~Vicente Vacas,
  %``Dynamically Generated Open and Hidden Charm Meson Systems,''
  Phys.\ Rev.\  D {\bf 76}, 074016 (2007)
%  [arXiv:hep-ph/0612179].
  %%CITATION = PHRVA,D76,074016;%%
%\cite{Molina:2008nh}
\bibitem{medio}
  R.~Molina, D.~Gamermann, E.~Oset and L.~Tolos,
  %``Charm and hidden charm scalar mesons in the nuclear medium,''
  Eur.\ Phys.\ J.\  A {\bf 42}, 31 (2009)
%  [arXiv:0806.3711 [nucl-th]].
  %%CITATION = EPHJA,A42,31;%%

%\cite{Faessler:2007gv}
\bibitem{Faessler:2007gv}
  A.~Faessler, T.~Gutsche, V.~E.~Lyubovitskij and Y.~L.~Ma,
  %``Strong and radiative decays of the Ds0*(2317) meson in the DK-molecule
  %picture,''
  Phys.\ Rev.\  D {\bf 76}, 014005 (2007)
%  [arXiv:0705.0254 [hep-ph]].
  %%CITATION = PHRVA,D76,014005;%%
%\cite{Kolomeitsev:2003ac}
\bibitem{lutz}
  E.~E.~Kolomeitsev and M.~F.~M.~Lutz,
  %``On heavy-light meson resonances and chiral symmetry,''
  Phys.\ Lett.\  B {\bf 582}, 39 (2004)
%  [arXiv:hep-ph/0307133].
  %%CITATION = PHLTA,B582,39;%%

%\cite{Guo:2006fu}
\bibitem{guo1}
  F.~K.~Guo, P.~N.~Shen, H.~C.~Chiang and R.~G.~Ping,
  %``Dynamically generated 0+ heavy mesons in a heavy chiral unitary
  %approach,''
  Phys.\ Lett.\  B {\bf 641}, 278 (2006)
%  [arXiv:hep-ph/0603072].
  %%CITATION = PHLTA,B641,278;%%
\bibitem{guo2}
  F.~K.~Guo, P.~N.~Shen and H.~C.~Chiang,
  %``Dynamically generated 1+ heavy mesons,''
  Phys.\ Lett.\  B {\bf 647}, 133 (2007)
%  [arXiv:hep-ph/0610008].
  %%CITATION = PHLTA,B647,133;%%
  %\cite{Guo:2006fu}
%\bibitem{guo3}
%  F.~K.~Guo, P.~N.~Shen, H.~C.~Chiang and R.~G.~Ping,
  %``Dynamically generated 0+ heavy mesons in a heavy chiral unitary
  %approach,''
%  Phys.\ Lett.\  B {\bf 641}, 278 (2006)
%  [arXiv:hep-ph/0603072].
  %%CITATION = PHLTA,B641,278;%%

%\cite{Gamermann:2007fi}
\bibitem{Gamermann2}
  D.~Gamermann and E.~Oset,
  %``Axial Resonances in the Open and Hidden Charm Sectors,''
  Eur.\ Phys.\ J.\  A {\bf 33}, 119 (2007)
%  [arXiv:0704.2314 [hep-ph]].
  %%CITATION = EPHJA,A33,119;%%
%\cite{Gamermann:2008jh}

%\cite{Faessler:2007us}
\bibitem{Faessler:2007us}
  A.~Faessler, T.~Gutsche, V.~E.~Lyubovitskij and Y.~L.~Ma,
  %``D* K molecular structure of the Ds1(2460) meson,''
  Phys.\ Rev.\  D {\bf 76}, 114008 (2007)
%  [arXiv:0709.3946 [hep-ph]].
  %%CITATION = PHRVA,D76,114008;%%
%\cite{Isgur:1989vq}
%\bibitem{isgur}
%  N.~Isgur and M.~B.~Wise,
  %``Weak Decays Of Heavy Mesons In The Static Quark Approximation,''
%  Phys.\ Lett.\  B {\bf 232}, 113 (1989).
  %%CITATION = PHLTA,B232,113;%%
%\cite{Isgur:1989ed}
%\bibitem{isgur1}
%  N.~Isgur and M.~B.~Wise,
  %``WEAK TRANSITION FORM-FACTORS BETWEEN HEAVY MESONS,''
%  Phys.\ Lett.\  B {\bf 237}, 527 (1990).
  %%CITATION = PHLTA,B237,527;%%
%\cite{Isgur:1991wq}
%\bibitem{isgur2}
 % N.~Isgur and M.~B.~Wise,
  %``Spectroscopy with heavy quark symmetry,''
 % Phys.\ Rev.\ Lett.\  {\bf 66}, 1130 (1991).
  %%CITATION = PRLTA,66,1130;%%
%\cite{Wise:1992hn}
\bibitem{wise}
  M.~B.~Wise,
  %``Chiral Perturbation Theory For Hadrons Containing A Heavy Quark,''
  Phys.\ Rev.\  D {\bf 45}, 2188 (1992).
  %%CITATION = PHRVA,D45,2188;%%
%\cite{Guo:2006rp}



\bibitem{raquel1}
  R.~Molina, D.~Nicmorus and E.~Oset,
  %``The \rho\rho interaction in the hidden gauge formalism and the f_0(1370)
  %and f_2(1270) resonances,''
  Phys.\ Rev.\  D {\bf 78}, 114018 (2008)
%  [arXiv:0809.2233 [hep-ph]].
  %%CITATION = PHRVA,D78,114018;%%
   %\cite{Amsler:2008zz}
\bibitem{pdg}
  C.~Amsler {\it et al.}  [Particle Data Group],
  %``Review of particle physics,''
  Phys.\ Lett.\  B {\bf 667}, 1 (2008).
  %%CITATION = PHLTA,B667,1;%%
%\cite{Geng:2008gx}
     %\cite{Klempt:2007cp}
\bibitem{klempt}
  E.~Klempt and A.~Zaitsev,
  %``Glueballs, Hybrids, Multiquarks. Experimental facts versus QCD inspired
  %concepts,''
  Phys.\ Rept.\  {\bf 454}, 1 (2007)
%  [arXiv:0708.4016 [hep-ph]].
  %%CITATION = PRPLC,454,1;%%


 %\cite{Crede:2008vw}
\bibitem{crede}
  V.~Crede and C.~A.~Meyer,
  %``The Experimental Status of Glueballs,''
  Prog.\ Part.\ Nucl.\ Phys.\  {\bf 63}, 74 (2009)
%  [arXiv:0812.0600 [hep-ex]].
  %%CITATION = PPNPD,63,74;%%
%\cite{Molina:2008jw}


\bibitem{geng}
  L.~S.~Geng and E.~Oset,
  %``Vector meson-vector meson interaction in a hidden gauge unitary approach,''
  Phys.\ Rev.\  D {\bf 79}, 074009 (2009)
%  [arXiv:0812.1199 [hep-ph]].
  %%CITATION = PHRVA,D79,074009;%%

%\cite{Nagahiro:2008um}
\bibitem{yamagata}
  H.~Nagahiro, J.~Yamagata-Sekihara, E.~Oset, S.~Hirenzaki and R. Molina,
  %``The $\gamma \gamma$ decay of the $f_0(1370)$ and $f_2(1270)$ resonances in
  %the hidden gauge formalism,''
  Phys.\ Rev.\  D {\bf 79}, 114023 (2009)
%  [arXiv:0809.3717 [hep-ph]].
  %%CITATION = PHRVA,D79,114023;%%
  
  %\cite{MartinezTorres:2009uk}
\bibitem{chinacola}
  A.~Martinez Torres, L.~S.~Geng, L.~R.~Dai, B.~X.~Sun, E.~Oset and B.~S.~Zou,
  %``Study of the $J/\psi \to \phi (\omega) f_2(1270)$, $J/\psi \to \phi
  %(\omega) f'_2(1525)$ and $J/\psi \to K^{*0}(892) \bar{K}^{* 0}_2(1430)$
  %decays,''
  Phys.\ Lett.\  B {\bf 680}, 310 (2009)
%  [arXiv:0906.2963 [nucl-th]].
  %%CITATION = PHLTA,B680,310;%%

\bibitem{chinavalgerman}  
%\cite{Geng:2009iw}
%\bibitem{Geng:2009iw}
  L.~S.~Geng, F.~K.~Guo, C.~Hanhart, R.~Molina, E.~Oset and B.~S.~Zou,
  %``Study of the $f_2(1270)$, $f_2'(1525)$, $f_0(1370)$ and $f_0(1710)$ in the
  %$J/\psi$ radiative decays,''
  arXiv:0910.5192 [hep-ph].
  %%CITATION = ARXIV:0910.5192;%%
%\cite{Branz:2009cv}
%\cite{Branz:2009cv}
\bibitem{BranzGeng}
  T.~Branz, L.~S.~Geng and E.~Oset,
  %``Two-photon and one photon-one vector meson decay widths of the $f_0(1370)$,
  %$f_2(1270)$, $f_0(1710)$, $f'_2(1525)$, and $K^*_2(1430)$,''
  Phys.\ Rev.\  D {\bf 81}, 054037 (2010)
  [arXiv:0911.0206 [Unknown]].
  %%CITATION = PHRVA,D81,054037;%%


  \bibitem{raquel2}
R. Molina, H. Nagahiro, A. Hosaka and E. Oset, 
Phys.\ Rev.\  D {\bf 80}, 014025 (2009)
%[arXiv:0903.3823v1 [hep-ph]].

%\cite{Molina:2009ct}
\bibitem{xyz}
  R.~Molina and E.~Oset,
  %``The Y(3940), Z(3930) and the X(4160) as dynamically generated resonances
  %from the vector-vector interaction,''
  Phys.\ Rev.\  D {\bf 80}, 114013 (2009)
  %\cite{Liang:2009sp}
\bibitem{weihong}
  W.~H.~Liang, R.~Molina and E.~Oset,
  %``Radiative open charm decay of the Y(3940), Z(3930), X(4160) resonances,''
  European Physical Journal A in print. 
  ArXiv:0912.4359 [hep-ph]. 
  %%CITATION = ARXIV:0912.4359;%%
  
%\cite{Branz:2010qw}
\bibitem{Branz:2010qw}
  T.~Branz, T.~Gutsche and V.~E.~Lyubovitskij,
  %``Possible hadronic molecule structure of the Y(3940) and Y(4140),''
  arXiv:1001.3959 [hep-ph].
  %%CITATION = ARXIV:1001.3959;%%

  %\cite{Bando:1984ej}
\bibitem{hidden1}
  M.~Bando, T.~Kugo, S.~Uehara, K.~Yamawaki and T.~Yanagida,
  %``Is Rho Meson A Dynamical Gauge Boson Of Hidden Local Symmetry?,''
  Phys.\ Rev.\ Lett.\  {\bf 54}, 1215 (1985).
  %%CITATION = PRLTA,54,1215;%%


%\cite{Bando:1987br}
\bibitem{hidden2}
  M.~Bando, T.~Kugo and K.~Yamawaki,
  %``Nonlinear Realization and Hidden Local Symmetries,''
  Phys.\ Rept.\  {\bf 164}, 217 (1988).
  %%CITATION = PRPLC,164,217;%%

 %\cite{Harada:2003jx}
\bibitem{hidden3}
  M.~Harada and K.~Yamawaki,
  %``Hidden local symmetry at loop: A new perspective of composite gauge boson
  %and chiral phase transition,''
  Phys.\ Rept.\  {\bf 381}, 1 (2003)
%  [arXiv:hep-ph/0302103].
  %%CITATION = PRPLC,381,1;%%
%\cite{Meissner:1987ge}
\bibitem{hidden4}
U.~G.~Meissner,
%``Low-Energy Hadron Physics From Effective Chiral Lagrangians With Vector
%Mesons,''
Phys.\ Rept.\  {\bf 161}, 213 (1988).
%%CITATION = PRPLC,161,213;%% 
%\cite{Bernard:1988db}


\bibitem{KSFR}
  Riazuddin and Fayyazuddin,
  %``Algebra of current components and decay widths of rho and K* mesons,''
  Phys.\ Rev.\  {\bf 147}, 1071 (1966).
  %%CITATION = PHRVA,147,1071;%%   
  
  \bibitem{sakurai} J.J. Sakurai, Currents and mesons (University of Chicago Press, Chicago Il
1969)
%\cite{Oller:1997ti}
\bibitem{Oller1}
  J.~A.~Oller and E.~Oset,
  %``Chiral Symmetry Amplitudes in the S-Wave Isoscalar and Isovector Channels
  %and the \sigma, f_0(980), a_0(980) Scalar Mesons,''
  Nucl.\ Phys.\  A {\bf 620}, 438 (1997)
  [Erratum-ibid.\  A {\bf 652}, 407 (1999)]
 % [arXiv:hep-ph/9702314].
  %%CITATION = NUPHA,A620,438;%%
%\cite{Oset:1997it}
\bibitem{OsetRa}
  E.~Oset and A.~Ramos,
  %``Non perturbative chiral approach to s-wave anti-K N interactions,''
  Nucl.\ Phys.\  A {\bf 635}, 99 (1998)
 % [arXiv:nucl-th/9711022].
  %%CITATION = NUPHA,A635,99;%%

 %\cite{Roca:2005nm}
\bibitem{Roca}
  L.~Roca, E.~Oset and J.~Singh,
  %``Low lying axial-vector mesons as dynamically generated resonances,''
  Phys.\ Rev.\  D {\bf 72}, 014002 (2005)
 % [arXiv:hep-ph/0503273].
 
 %\cite{Nagahiro:2008cv}
\bibitem{hidekoroca}
  H.~Nagahiro, L.~Roca, A.~Hosaka and E.~Oset,
  %``Hidden gauge formalism for the radiative decays of axial-vector mesons,''
  Phys.\ Rev.\  D {\bf 79} (2009) 014015
%  [arXiv:0809.0943 [hep-ph]].

%\cite{Titov:2000bn}
\bibitem{Titov:2000bn}
  A.~I.~Titov, B.~Kampfer and B.~L.~Reznik,
  %``Production of Phi mesons in near-threshold pi N and N N reactions,''
  Eur.\ Phys.\ J.\  A {\bf 7}, 543 (2000).
  %[arXiv:nucl-th/0001027].
  %%CITATION = EPHJA,A7,543;%%

  %\cite{Titov:2001yw}
\bibitem{Titov:2001yw}
  A.~I.~Titov, B.~Kampfer and B.~L.~Reznik,
  %``Production of omega and Phi mesons in near-threshold pi N reactions:
  %Baryon resonances and validity of the OZI rule,''
  Phys.\ Rev.\  C {\bf 65}, 065202 (2002).
  %[arXiv:nucl-th/0102032].
  %%CITATION = PHRVA,C65,065202;%%
  
   \bibitem{Navarra}
  F.~S.~Navarra, M.~Nielsen and M.~E.~Bracco,
  %``D* D pi form factor revisited,''
  Phys.\ Rev.\  D {\bf 65} 037502 (2002) 
%  [arXiv:hep-ph/0109188].
  %%CITATION = PHRVA,D65,037502;%% 
  
   %\cite{Ahmed:2001xc}
\bibitem{Ahmed}
  S.~Ahmed {\it et al.}  [CLEO Collaboration],
  %``First measurement of Gamma(D*+),''
  Phys.\ Rev.\ Lett.\  {\bf 87}, 251801 (2001)
%  [arXiv:hep-ex/0108013].
  %%CITATION = PRLTA,87,251801;%%


 %\cite{Belyaev:1994zk}
\bibitem{Belyaev}
  V.~M.~Belyaev, V.~M.~Braun, A.~Khodjamirian and R.~Ruckl,
  %``D* D pi and B* B pi couplings in QCD,''
  Phys.\ Rev.\  D {\bf 51}, 6177 (1995)
%  [arXiv:hep-ph/9410280].
  %%CITATION = PHRVA,D51,6177;%%
 %\cite{Colangelo:1994es}
 
\bibitem{Colangelo}
  P.~Colangelo, G.~Nardulli, A.~Deandrea, N.~Di Bartolomeo, R.~Gatto and F.~Feruglio,
  %``On the coupling of heavy mesons to pions in QCD,''
  Phys.\ Lett.\  B {\bf 339}, 151 (1994)
 % [arXiv:hep-ph/9406295].
  %%CITATION = PHLTA,B339,151;%%

  %%CITATION = PHRVA,D79,014015;%%
  %%CITATION = PHRVA,D72,014002;%%
%\bibitem{Isgur}
%  N.~Isgur and M.~B.~Wise,
  %``Spectroscopy with heavy quark symmetry,''
%  Phys.\ Rev.\ Lett.\  {\bf 66}, 1130 (1991).
  %%CITATION = PRLTA,66,1130;%%
  
  %\cite{Di Pierro:2001uu}

%\cite{Matsuki:1997da}
%\bibitem{Matsuki}
 % T.~Matsuki and T.~Morii,
  %``Spectroscopy of heavy mesons expanded in 1/m(Q),''
%  Phys.\ Rev.\  D {\bf 56}, 5646 (1997)
%  [Austral.\ J.\ Phys.\  {\bf 50}, 163 (1997)]
%  [arXiv:hep-ph/9702366].
  %%CITATION = AUJPA,50,163;%%



\end{thebibliography}
\end{document}